\documentclass{emulateapj}
\usepackage{natbib}
\begin{document}
\title{The Michigan/MIKE Fiber System Survey of Stellar Radial Velocities in Dwarf Spheroidal Galaxies: Acquisition and Reduction of Data\footnote{This paper includes data obtained with the 6.5 meter Magellan Telescopes located at Las Campanas Observatory, Chile.}}
\shorttitle{Magellan/MMFS Survey of dSph Kinematics}
\author{Matthew G. Walker\altaffilmark{1}, Mario Mateo\altaffilmark{1}, Edward W. Olszewski\altaffilmark{2}, Rebecca Bernstein\altaffilmark{1}, Bodhisattva Sen\altaffilmark{3}, and Michael Woodroofe\altaffilmark{3}}
\altaffiltext{1}{Department of Astronomy, University of Michigan, 830 Dennison Building, Ann Arbor, MI 48109-1042}
\altaffiltext{2}{Steward Observatory, The University of Arizona, Tucson, AZ 85721}
\altaffiltext{3}{Department of Statistics, University of Michigan, 439 West Hall, Ann Arbor, MI 48109-1107}

\begin{abstract} 
We introduce a stellar velocity survey of dwarf spheroidal galaxies, undertaken using the Michigan/MIKE Fiber System (MMFS) at the Magellan/Clay 6.5 m telescope at Las Campanas Observatory.  As of 2006 November we have used MMFS to collect 6415 high-resolution ($R= 20000-25000$) spectra from 5180 stars in four dwarf spheroidal galaxies: Carina, Fornax, Sculptor and Sextans.  Spectra sample the range $5140-5180$ \AA, which includes the prominent magnesium triplet absorption feature.  We measure radial velocity (RV) to a median precision of 2.0 km s$^{-1}$ for stars as faint as $V \sim 20.5$.  From the spectra we also are able to measure the strength of iron and magnesium absorption features using spectral indices that correlate with effective temperature, surface gravity and chemical abundance.  Measurement of line strength allows us to identify interloping foreground stars independently of velocity, and to examine the metallicity distribution among dSph members.  Here we present detailed descriptions of MMFS, our target selection and spectroscopic observations, the data reduction procedure, and error analysis.  We compare our RV results to previously published measurements for individual stars.  In some cases we find evidence for a mild, velocity-dependent offset between the RVs we measure using the magnesium triplet and previously published RV measurements derived from the infrared calcium triplet.  In companion papers we will present the complete data sets and kinematic analyses of these new observations.  
\end{abstract}

\section{Introduction}

The dwarf spheroidal (dSph) satellite companions of the Milky Way (MW) contain a wealth of information regarding the formation and ongoing evolution of the Local Group.  The $\sim 15$ known MW dSphs are the nearest, smallest, and faintest known galaxies ($0.5 \leq$ R/kpc $\leq 1$; $10^4 \leq$ L/L${\sun}\leq 10^7$; $20 \leq$ D/kpc $\leq 250$; \citealt{mateo98} and references therein; \citealt{willman05a,belokurov06a}).  From their resolved stellar populations one can study in detail dSph morphologies, star formation histories, chemical abundances, and kinematics.  Given their surface brightness profiles, the stellar velocity dispersions ($\sigma_v \sim 10$ km s$^{-1}$) of these galaxies indicate a larger mass than could be contributed by any reasonable equilibrium stellar population.  Applying the virial theorem, one derives for the MW dSphs masses of $\sim 10^7-10^8$ M$_{\sun}$ and mass-to-light (M/L) ratios ranging from five to several hundred (\citealt{mateo98} and references therein).  To the extent that the virial theorem is applicable despite the tidal field exerted by the Milky Way (\citealt{pp95,oh95}), the derived M/L ratios place at least some dSphs among the most dark-matter dominated galaxies known, generally with negligible baryonic contribution to their total mass.  In this sense the MW dwarfs offer an ideal testbed with which to constrain the behavior of dark matter at perhaps the most difficult scale to model---the lower extremum of the halo mass function.  

Efforts to characterize the internal kinematics of dSphs have expanded in scope over the past quarter-century.  In a pioneering study, \citet{aaronson83} considered the possible dark matter (DM) content of dSphs based on the radial velocity (RV) dispersion of just three Draco stars.  The need to invoke DM to describe dSph kinematics was placed on firmer footing after samples of tens of stars \citep{aaronson87,mateo91,mateo93,suntzeff93,edo95,queloz95,hargreaves94,hargreaves96b,vogt95,mateo98b} confirmed Aaronson's result.  RV samples for nearly 100 stars accumulated over several epochs \citep{armandroff95,edo96,hargreaves96a,hargreaves96b} showed that the magnitudes of dSph velocity dispersions were not unduly influenced by the presence of binary stars or motion within the stellar atmospheres.  Recent RV samples for more than 100 stars per dSph \citep{kleyna01,kleyna02,kleyna03,kleyna04,tolstoy04,munoz06,battaglia06,walker06a,walker06b,koch06b}, yield the general result that dSph velocity profiles appear isothermal over most of the visible faces of dSphs, though there is some dispute over the velocity behavior near the nominal tidal radius \citep{kleyna04,munoz05}.  Moreover, the growing number of measurements for stars at and even well beyond the nominal tidal radius (e.g., \citealt{munoz06}; \citealt{sohn06}; Mateo et al. in preparation) has re-energized debate about the degree to which external tidal forces influence dSph stellar kinematics.   

Studies by \citet{merritt97,wilkinson02,wang05} have shown theoretically that velocity samples for $\sim 1000$ stars are capable of distinguishing among spherical mass models while relaxing some of the assumptions (e.g., constant mass-to-light ratio and velocity isotropy) characteristic of early analyses.  The present observational challenge thus is to acquire data sets consisting of $\ge 1000$ stellar RVs with spatial sampling that extends over the entire face of the dSph.  In order to resolve narrow dSph velocity distributions one must work at high resolution.  Stellar targets sufficiently bright (V$\leq 21$) for measuring radial velocity and chemical abundances are generally confined to the dSph red giant branch.  The development of multi-object fiber and/or slit-mask spectrographs with echelle resolution has fueled a recent boom in this field.  Recent and ongoing work using VLT/FLAMES/GIRAFFE \citep{tolstoy04,battaglia06,munoz06}, WHT/WYFFOS \citep{kleyna04}, Keck/DEIMOS \citep{sohn06,koch06b}, MMT/Hectochelle (Mateo et al.\ in preparation) has yielded RV samples for hundreds of stars in several MW dSphs.  These studies have produced several results that challenge previous assumptions.  \citet{tolstoy04} find evidence for multiple stellar populations in Sculptor that follow distinct distributions in position, metallicity, \textit{and} RV.  \citet{battaglia06} produce a similar result in a study of Fornax.  \citet{munoz06} discover several widely separated Carina members, which they interpret as evidence of tidal streaming; they also attribute a secondary clump in Carina's velocity distribution to contamination by stars once associated with the Large Magellanic Cloud.  Mateo et al.\ (in preparation) and Sohn et al.\ (2006) detect a kinematic signature of tidal streaming from the outer regions of Leo I.  It is becoming clear that dSphs are complex systems that can be characterized adequately only with data sets of the sort coming from large surveys.

Here we introduce an independent dSph kinematic survey undertaken with a new instrument, the Michigan-MIKE Fiber System (MMFS).  MMFS provides multi-object observing capability using the dual-channel Magellan Inamori Kyocera Echelle (MIKE) spectrograph at the Magellan 6.5m Clay Telescope.  Ours is the only large-scale dSph survey to measure RV using primarily the Mg-triplet ($\lambda \sim 5170$ \AA) absorption feature; others use the infrared calcium triplet near 8500 \AA.   Thus far we have used MMFS to measure 6415 RVs to a median precision of $\pm 2.0$ km s$^{-1}$ for 5180 stars in four MW dSphs.  Of the measured stars, which span the magnitude range $17 \le V \le 20.5$, approximately 470 are members of the Carina dSph, 1900 are members of Fornax, 990 are members of Sculptor, and 400 are members of Sextans.  The remaining 1400 measured stars are likely dwarfs contributed by the MW foreground.  In addition, the MMFS spectra are of sufficient quality that we can measure magnesium line strengths via a set of spectral indices that correlate with stellar parameters such as effective temperature and surface gravity, and which trace chemical abundances of iron-peak and alpha elements.  The spectral indices provide independent constraints on dSph membership and allow us to probe the chemical abundance distribution within a given dSph.

Our focus in this paper is to provide a thorough description of MMFS, the observations, data reduction procedure, derivation of measurement uncertainties, and comparison with previous velocity measurements.  This material will be referenced in companion papers (Walker et al.\ in preparation) in which we present all data and analyses. 

\section{The Michigan-MIKE Fiber System}
\label{sec:mmfs}

In its standard slit observing mode the MIKE spectrograph delivers high-resolution spectra spanning visible and near-infrared wavelengths.  MIKE's first optical element is a coated glass dichroic that reflects/transmits incoming light into one of two independent channels.  The ``blue'' channel has wavelength coverage 3200-5000 \AA, while the ``red'' channel covers 4900-10000 \AA.  The dual-beam design allows for independent optimization of throughput and dispersion characteristics over each range of spectrum (see \citet{bernstein03} for a complete description of MIKE).  

For multi-object observations using Magellan+MMFS, MIKE is modified from its standard slit configuration to receive light from up to 256 fibers.  In fiber mode MIKE is backed $2$ m from the Nasmyth port, where it remains fixed with respect to gravity.  Inserted at the port is a conical structure onto which an (interchangeable) aluminum plug plate is mounted.  The plug plate holds fibers against a telecentrator lens that ensures the 1.4-arcsec fiber apertures accept light parallel to the optical axis and along the focal surface, regardless of position in the target field.  A drum lens at the entrance to each fiber forms an image of the pupil on the the $175$-$\mu$m fiber core.  The drum lens and the natural focal ratio degradation of the fiber convert the incoming beam from F/11 to F/3.5 (more similar to MIKE's camera optics).  Fibers run from the plug plate through a junction box, where they are bundled and sorted into one of two 128-fiber assemblies that enter the two channels of the spectrograph.  Light losses are minimized by the short fiber length; from end to end,  blue (red) fibers extend just 2.43 m (2.29 m).  

At the spectrograph end, the two 128-fiber assemblies replace MIKE's standard injection optics.  Due to the spectrograph's fixed optical elements, blue and red channels remain optimized for smaller and larger wavelengths, respectively.  In order to fit spectra from all 128 fibers onto a single detector in each channel, narrow-band filters are used to isolate a single order of spectrum for each target object.  We use filters in both channels that isolate the spectral region $5130-5185$ \AA.  In late-type stellar spectra this region includes the prominent MgI triplet (MgT) absorption feature and an assortment of FeI and NiI absorption lines.  The MgT filter yields useful spectra through either MIKE channel, though the blue channel offers higher spectral resolution ($\sim 0.059$ \AA\ pix$^{-1}$; $R \sim 25000$) and greater throughput than does the red channel ($\sim 0.073$ \AA\ pix$^{-1}$; $R\sim 20000$).  MMFS thus has the ability to acquire high-resolution spectra from up to 256 objects simultaneously.  The usable field of view subtends an angular diameter of $20$ arcmin.  

\section{Observations}
\label{sec:observations}
\subsection{Target Selection}

For our survey, MMFS target star candidates are selected from V,I imaging data taken prior to spectroscopic observing runs.  We select MMFS target candidates from the dSph red giant branch (RGB) of the resulting color-magnitude diagram (CMD).  We expect a fraction of the selected candidates to be foreground Milky Way dwarfs with magnitudes and colors placing them on or near the dSph RGB.  The fraction of interlopers naturally increases for lines of sight along which the dSph surface density is relatively small.  The efficiency with which our CMD-based selection identifies bona fide dSph red giants thus varies from galaxy to galaxy and with position within each galaxy, and foreground stars must subsequently be identified and removed from kinematic samples.  

Details of the photometry used to select red giant candidates in the Fornax and Sculptor dSphs are published elsewhere.  Fornax candidates were selected using the V,I photometric data described in \citet{walker06a}.  Sculptor candidates were selected using the V,I photometry of \citet{coleman05a}, who generously provided their data set.

\subsubsection{Sextans}

For the Sextans dSph, red giant candidates were chosen based on V,I photometry taken with the MDM 2.4 m Hiltner Telescope and 8k Mosaic CCD detector.  Observations took place the nights of 2004 February 11-17.  Three of the seven nights were photometric; the remaining nights had mixed observing conditions.  We observed 25 Sextans fields, each with dimensions $24\arcmin \times 24\arcmin$, forming a square array with adjacent field centers separated by $20\arcmin$.  We exposed for 360s in I and 540s in V.  Standard reduction steps included overscan, bias, and flat-field corrections using twilight sky flats.  We used the two-dimensional stellar photometry program DoPHOT \citep{schechter93} to identify stars and measure instrumental magnitudes.  We placed the photometry in the Kron-Cousins system \citep{bessell76} using 50 photometric standard stars \citep{landolt92} observed during the run.  We used the $4\arcmin$ overlap between adjacent fields to remove photometric zero-point offsets.  Formal error values returned by DoPHOT and multiple measurements from overlapping fields indicate the typical photometric accuracy is $\pm 0.04$ mag. 

\subsubsection{Carina}

Over several MMFS observing runs, red giant candidates in the Carina dSph were chosen using three independent photometric data sets.  Carina targets observed with MMFS in 2004 March were selected from B,R photometry obtained with the CTIO 4m telescope with Mosaic CCD detector during 1999 September.  Carina targets observed with MMFS in 2005 February were chosen in part from the same B,R mosaic photometry, but targets in outer fields were selected from V,I photometry taken by Kaspar von Braun at the 2.4 m Swope Telescope (Las Campanas Observatory) in 2004 December.  Carina targets observed with MMFS in 2006 March were selected from V,I photometry taken by Patrick Seitzer with the CTIO 0.9 m Schmidt Telescope in 2005 December.  We used DoPHOT to reduce the photometry in each case.  

In order to place the three photometric data sets for Carina on a common V,I system, first we used 122 photometric standard stars in the Carina field of \citet{stetson00} to apply zero-point corrections to the Schmidt instrumental V,I magnitudes.  Residuals after applying these shifts have rms values 0.033 mag in V and 0.035 mag in I.  We followed the same procedure in calibrating the 2004 V,I photometry (residuals have rms 0.046 mag in V and 0.048 mag in I).  Next we used a piecewise linear transformation to place the B,R mosaic photometry on the Stetson V,I system.  Coefficients were determined using 286 Stetson standard stars detected in the mosaic photometry.  Over the color range $0.6 \leq$ V-I $\leq 1.8$, residuals have rms $\sim 0.03$ mag for the V and V-I transformations.  We monitored for consistency by comparing the calibration results for stars detected in both the mosaic and Schmidt data.  For 637 stars present in both data sets, flux deviations have mean values $\langle V_{schmidt}-V_{mosaic} \rangle = 0.011$ mag and $\langle I_{schmidt}-I_{mosaic} \rangle = 0.003$ mag, with rms values of 0.072 mag and 0.088 mag, respectively.  These values are reasonable considering the undersampling and blending in the Schmidt data.

\subsubsection{Globular Clusters}

For the purpose of calibrating spectral-line index/abundance relations, we have also observed stellar targets in seven Galactic globular clusters.  For clusters NGC 104 (47 Tucana), NGC 288, and NGC 7099 (Messier 30) we chose targets on the basis of V,I photometry obtained in 2005 July with the 40-inch telescope at Las Campanas Observatory, Chile.  For clusters NGC 5927, NGC 6121 (Messier 4), NGC 6171 (Messier 107), and NGC 6397 we selected targets using V,I photometry obtained in 2005 June with the Magellan/Baade 6.5m Telescope at Las Campanas.  For all clusters, instrumental magnitudes were measured with DoPHOT and then placed on the V,I system using stars in common with Stetson standard fields.  For NGC 6171, which is absent from the Stetson catalog, we applied the same transformations determined for NGC 5927, a cluster observed on the same night and at similar airmass.  Direct comparison to Stetson stars indicates that the globular cluster photometry is accurate to within $\sim 0.04-0.08$ magnitudes.

\subsection{CMDs and maps for dSph targets}

\begin{figure*}
  \epsscale{1}
 \plotone{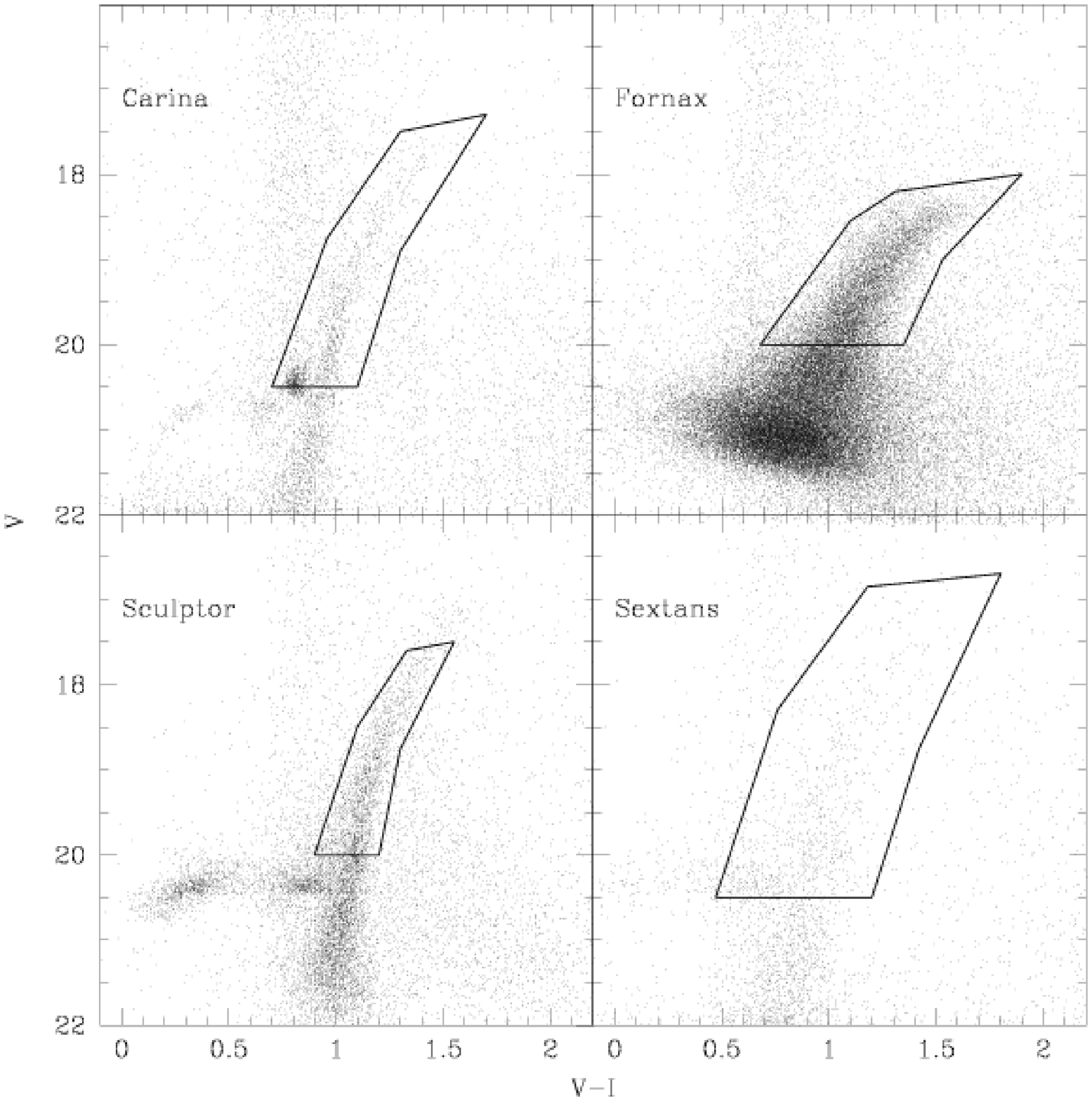}
  \caption{ V,I color-magnitude diagrams for the four dSphs observed with MMFS.  Stars enclosed by polygons are considered red giant candidates and eligible MMFS targets}
  \label{fig:cmd}
\end{figure*}

\begin{figure*}
  \epsscale{1}
 \plotone{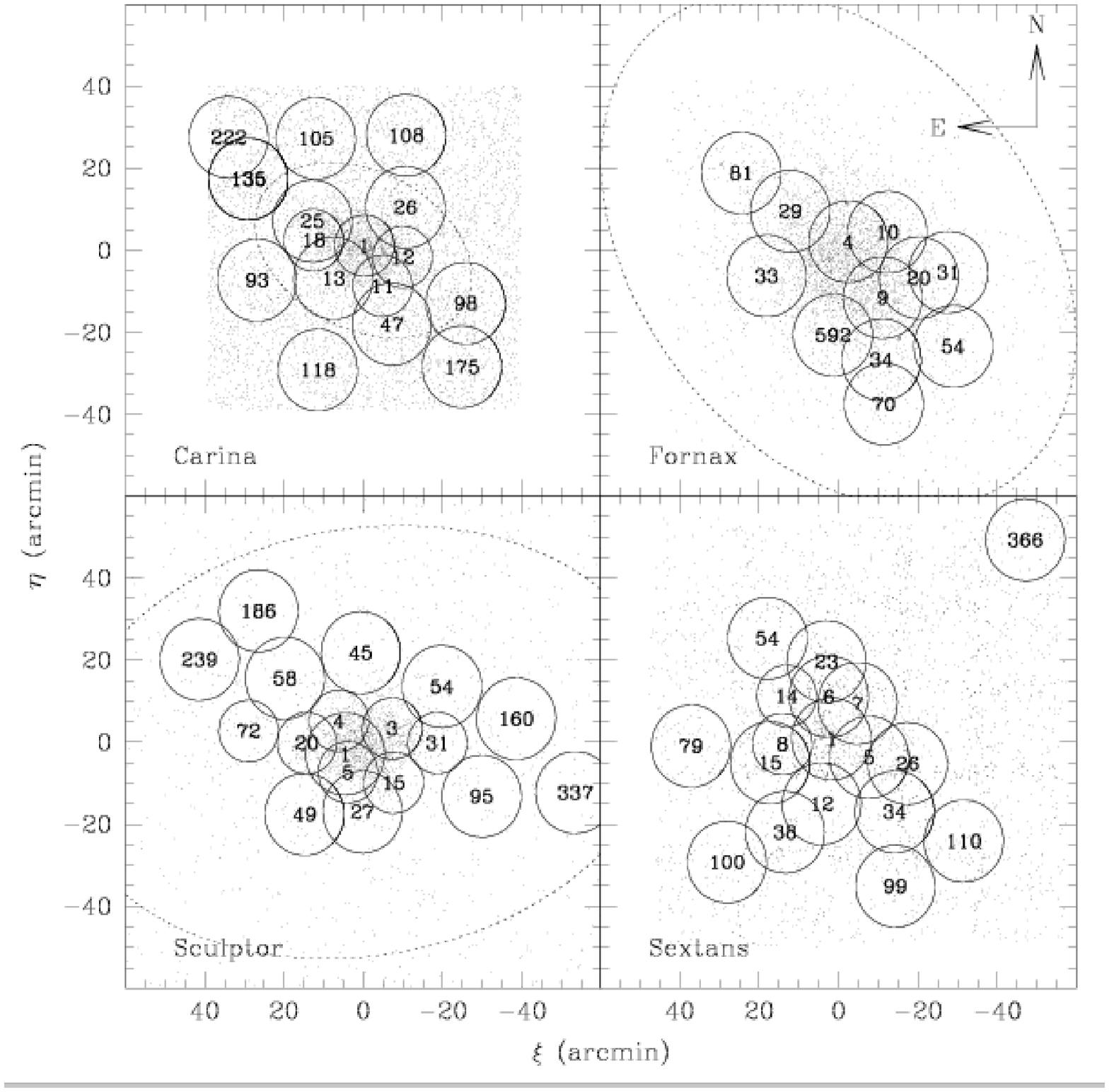}
  \caption{  Maps of red giant candidates identified in Figure \ref{fig:cmd}.  Circles indicate fields observed with MMFS (field number corresponds to values in Column 2 of Table \ref{tab:observations}.  Dotted ellipses correspond to (nominal) tidal radii, as identified by \citet{king62} model fits to the surface brightness profiles by \citet{ih95}.  For Sextans the tidal radius ($r_t \sim 160\arcmin$) lies beyond the plotted region.  All sky positions are given in standard coordinates with origin at the dSph center.  We observed three distinct sets of target stars in each of the densely populated Fornax fields 4 and 9.}    
 \label{fig:selection_map}
\end{figure*}

Figures \ref{fig:cmd} plots CMDs for Carina, Fornax, Sculptor and Sextans.  Polygons enclose stars we consider to be red-giant candidates, and therefore eligible MMFS targets.  For Carina, Sculptor and Sextans the selection region extends from near the tip of the red giant branch to the horizontal branch at $V \sim 20.5$.  For Fornax, the greater stellar density allows us to increase efficiency by selecting from only those red-giant candidates having $V < 20$.  Figure \ref{fig:selection_map} maps the locations of red giant candidates in each dSph.  Circles in Figure \ref{fig:selection_map} indicate MMFS fields that have been observed as of 2006 November.  Initial MMFS runs gave priority to target-rich central fields, with goals of building sample size and evaluating instrument performance.  Later runs gave preference to fields along the morphological major axis and/or at large radius.  The partial overlap of many adjacent fields allows us to quantify velocity variabilty using repeat measurements.  

The $1.4\arcsec$ fiber apertures require that target coordinates be accurate to within $\sim 0.2\arcsec$.  DoPHOT's centroid algorithm returns the (x,y) CCD position of each identified star.  We convert these to equatorial coordinates using the IRAF routines TFINDER and CCTRANS.  We then tie the astrometry to the USNO-B1 system by identifying up to several hundred USNO stars per CCD frame.  From the derived coordinates of the same stars in overlapping fields we find a typical rms scatter of 0.24\arcsec.


\subsection{Spectroscopic Observing Procedure}

In a typical night we take $30$-$50$ zero-second exposures just prior to evening twilight.   During twilight we take $3$-$5$ exposures of the scattered solar spectrum, to allow for wavelength calibration and provide a velocity zero point, followed by $3 \times 60$s quartz lamp exposures and a $1 \times 300$s Th-Ar arc-lamp exposure.

Under optimal observing conditions we take $3 \times 2400$s exposures of target fields, immediately followed by $3 \times 60$s flat field (quartz) exposures and $1 \times 300$s Th-Ar.  Other observations taken during the night include single-fiber, $1 \times 30$s exposures of radial velocity standard stars, immediately followed by quartz and Th-Ar calibration exposures taken in the usual manner.  The procedure allows for observation of 2-4 dSph fields per night.

Table \ref{tab:observations} logs all MMFS observations taken for dSph and globular cluster fields as of 2006 November.  The first two columns identify the field by galaxy name and field number (observed fields are mapped by field number in Figure \ref{fig:selection_map}).  Columns 3-7 in Table \ref{tab:observations} list, respectively, the heliocentric Julian date at the midpoint of the first exposure, UT date at the midpoint of the first exposure, total exposure time, number of red giant candidates to which we assigned fibers, and the number of these for which the radial velocity was measured succesfully.  The final column identifies repeat observations of fields for which we judged during the course of the run that the initial observation was insufficient.  

\begin{deluxetable*}{lccccccl}
  \tabletypesize{\scriptsize}
  \tablewidth{0pc}
  \tablecaption{ MMFS observations of dSph and globular cluster fields as of 2006 November}
  \tablehead{\colhead{Galaxy}&\colhead{Field}&\colhead{HJD $-2.45 \times 10^6$}&\colhead{UT Date}&\colhead{Exposure Time }&\colhead{Targets}&\colhead{Velocities}&\colhead{Notes}\\
    \colhead{}&\colhead{}&\colhead{(days)}&\colhead{}&\colhead{(s)}&\colhead{}&\colhead{}&\colhead{}
  }
  \startdata
Sextans&8&3086.781&2004 Mar. 22&$2 \times 1800$&120&32\\
Carina&18&3087.533&2004 Mar. 23&$4 \times 2400$&173&83\\
Sextans&8&3087.716&2004 Mar. 23&$4 \times 2400$&120&&repeat from 2004 Mar. 22\\
Carina&1&3088.540&2004 Mar. 24&$3 \times 2400$&219&92\\
Sextans&14&3088.704&2004 Mar. 24&$4 \times 2400$&139&49\\
Carina&1&3089.540&2004 Mar. 25&$4 \times 2400$&220&&repeat from 2004 Mar. 24\\
Sextans&1&3089.724&2004 Mar. 25&$2 \times 2400$&196&66\\
&&&&$1 \times 3000$\\
Carina&12&3090.551&2004 Mar. 26&$4 \times 2400$&223&77\\
Sextans&7&3090.704&2004 Mar. 26&$2 \times 2700$&105&48\\
&&&&$1\times 3600$\\
Carina&11&3091.520&2004 Mar. 27&$2 \times 3000$&190&115\\
&&&&$1 \times 3600$\\
Sextans&5&3091.698&2004 Mar. 27&$1 \times 3000$&140&50\\
&&&&$1 \times 3600$\\
Carina&1&3092.519&2004 Mar. 28&$2 \times 3000$&220&&repeat from 2004 Mar. 24, Mar. 25\\
&&&&$1 \times 3600$\\
Sextans&1&3092.704&2004 Mar. 28&$1 \times 2400$&200&&repeat from 2004 Mar. 25\\
&&&&$1 \times 3000$\\
&&&&$1 \times 3600$\\
Carina&26&3093.531&2004 Mar. 29&$1 \times 2700$&85&21\\
&&&&$1 \times 3000$\\
\\
\hline
\\
Sculptor&20&3286.629&2004 Oct. 8&$4 \times 1800$&135&110\\
Fornax&15&3286.755&2004 Oct. 8&$1 \times 1800$&220&191\\
&&&&$1 \times 2400$\\
&&&&$1 \times 2700$\\
&&&&$1 \times 3600$\\
Sculptor&3&3287.552&2004 Oct. 9&$2 \times 1800$&224&168\\
&&&&$1 \times 2400$\\
&&&&$1 \times 2600$\\
Sculptor&31&3287.694&2004 Oct. 9&$2 \times 1800$&87&69\\
&&&&$2 \times 2400$\\
Fornax&15&3287.826&2004 Oct. 9&$1 \times 2100$&222&&repeat from 2004 Oct. 8\\
&&&&$1 \times 2400$\\
&&&&$1 \times 2200$\\
Sculptor&4&3288.571&2004 Oct. 10&$2 \times 2400$&224&196\\
&&&&$1 \times 2500$\\
Sculptor&72&3288.679&2004 Oct. 10&$2 \times 2400$&31&19\\
&&&&$1 \times 2100$\\
Fornax&15&3288.801&2004 Oct. 10&$1 \times 2900$&222&&repeat from 2004 Oct. 8,9\\
&&&&$1 \times 3000$\\
Sculptor&3&3289.525&2004 Oct. 11&$1 \times 3600$&224&&repeat from 2004 Oct. 9\\ 
Sculptor&5&3289.593&2004 Oct. 11&$3 \times 2400$&224&176\\
Sculptor&15&3289.712&2004 Oct. 11&$3 \times 2400$&138&128\\
&&&&$1 \times 2700$\\
Sculptor&20&3289.868&2004 Oct. 11&$1 \times 3000$&135&&repeat from 2004 Oct. 8\\
\\
\hline
\\
Sextans&12&3408.709&2005 Feb. 7&$4 \times 2400$&75&34\\
Carina&47&3409.650&2005 Feb. 8&$3 \times 2400$&129&109\\
Sextans&34&3409.774&2005 Feb. 8&$3 \times 2400$&83&75\\
Carina&13&3410.547&2005 Feb. 9&$4 \times 1800$&223&139\\
Carina&108&3410.668&2005 Feb. 9&$2 \times 1800$&95&73\\
Sextans&7&3410.758&2005 Feb. 9&$5 \times 2400$&185&127\\
Carina&25&3411.553&2005 Feb. 10&$6 \times 2400$&223&148\\
Sextans&1&3411.776&2005 Feb. 10&$4 \times 2700$&223&113\\
Sextans&366&3412.725&2005 Feb. 11&$3 \times 2700$&44&27\\
Sextans&38&3412.846&2005 Feb. 11&$2 \times 2500$&52&38\\
Carina&25&3415.550&2005 Feb. 14&$4 \times 1800$&224&&repeat from 2005 Feb. 10\\
Carina&108&3415.678&2005 Feb. 14&$3 \times 1800$&95&&repeat from 2005 Feb. 9\\
Sextans&1&3415.787&1005 Feb. 14&$4 \times 2400$&224&&repeat from 2005 Feb. 10\\
Carina&118&3416.555&2005 Feb. 15&$3 \times 2400$&119&90\\
Sextans&5&3416.683&2005 Feb. 15&$4 \times 2400$&177&117\\
Sextans&26&3416.837&2005 Feb. 15&$1 \times 1800$&118&61\\
Carina&13&3417.544&2005 Feb. 16&$2 \times 2400$&224&&repeat from 2005 Feb. 9\\
Carina&222&3417.630&2005 Feb. 16&$3 \times 2400$&81&72\\
Sextans&15&3417.761&2005 Feb. 16&$4 \times 2400$&119&91\\
Carina&26&3418.552&2005 Feb. 17&$4 \times 2400$&80&48\\
Sextans&23&3418.704&2005 Feb. 17&$3 \times 2400$&104&72\\
\\
\hline
\\
NGC 7099&1&3663.533&2005 Oct. 20&$3 \times 900$&59&58&\\
47 Tuc&1&3663.584&2005 Oct. 20&$3 \times 600$&92&71&\\
&&&&$1 \times 300$\\
Sculptor&49&3663.657&2005 Oct. 20&$3 \times 2400$&57&27\\
Sculptor&45&3663.763&2005 Oct. 20&$3 \times 2400$&52&32\\
NGC 7099&1&3664.492&2005 Oct. 20&$4 \times 300$&59&38&\\
47 Tuc&1&3664.510&2005 Oct. 21&$3 \times 300$&92&78&\\
Sculptor&239&3664.558&2005 Oct. 21&$3 \times 2400$&26&13\\
Sculptor&95&3664.656&2005 Oct. 21&$4 \times 1800$&33&24\\
Fornax&10&3664.771&2005 Oct. 21&$3 \times 1800$&223&201\\
&&&&$2 \times 2100$\\
NGC 288&1&3665.513&2005 Oct. 22&$4 \times 300$&63&40&\\
Sculptor&58&3665.559&2005 Oct. 22&$3 \times 2400$&41&20\\
Sculptor&27&3665.662&2005 Oct. 22&$3 \times 2400$&110&87\\
Fornax&29&3665.772&2005 Oct. 22&$4 \times 2400$&216&177\\
NGC 288&1&3666.504&2005 Oct. 22&$4 \times 300$&64&63&\\
Sculptor&54&3666.546&2005 Oct. 23&$3 \times 2400$&65&48\\
Fornax&34&3666.655&2005 Oct. 23&$4 \times 2400$&163&147\\
Fornax&592&3666.796&2005 Oct. 23&$2 \times 2700$&127&109\\
&&&&$1 \times 2400$\\
\\
\hline
\\
Carina&135&3800.581&2006 Mar. 6&$4 \times 2400$&202&47\\
Sextans&110&3800.762&2006 Mar. 6&$3 \times 2400$&77&43\\
Carina&175&3801.535&2006 Mar. 7&$4 \times 2400$&195&103\\
Sextans&99&3801.696&2006 Mar. 7&$3 \times 2400$&74&51\\
Sextans&79&3801.820&2006 Mar. 7&$1 \times 2400$&69&42\\
NGC 5927&1&3801.8611&2006 Mar. 7&$3 \times 300$&111&111&\\
Carina&175&3802.542&2006 Mar. 8&$3 \times 2400$&195&&repeat from 2006 Mar. 7\\
Sextans&99&3802.656&2006 Mar. 8&$3 \times 2400$&74&&repeat from 2006 Mar. 7\\
Sextans&79&3802.764&2006 Mar. 8&$3 \times 2400$&69&&repeat from 2006 Mar. 7\\
M4&1&3802.858&2006 Mar. 8&$3 \times 300$&64&64&\\
NGC 6397&1&3802.894&2006 Mar. 8&$2 \times 150$&31&31&\\
&&&&$1 \times 300$\\
Carina&98&3803.601&2006 Mar. 9&$3 \times 2400$&210&75\\
Sextans&54&3803.716&2006 Mar. 9&$3 \times 2400$&59&32\\
NGC 6171&1&3803.861&2006 Mar. 9&$3 \times 300$&61&55&\\
Carina&105&3804.524&2006 Mar. 10&$4 \times 2400$&166&45\\
Sextans&6&3804.673&2006 Mar. 10&$6 \times 2400$&197&61\\
Carina&135&3805.515&2006 Mar. 10&$2 \times 2400$&202&&repeat from 2006 Mar. 6\\
Carina&93&3805.597&2006 Mar. 11&$4 \times 2400$&166&90\\
Sextans&100&3805.736&2006 Mar. 11&$4 \times 2400$&59&22\\
\\
\hline
\\
Sculptor&1&4017.5547&2006 Oct. 9&$4 \times 2700$&223&186\\
Sculptor&160&4018.5527&2006 Oct. 10&$3 \times 2700$&28&21\\
Sculptor&186&4018.6665&2006 Oct. 10&$3 \times 2700$&25&10\\
Sculptor&337&4019.5623&2006 Oct. 11&$3 \times 2400$&18&11\\
Fornax&4.1&4019.6829&2006 Oct. 11&$3 \times 2400$&224&208\\
Fornax&9.1&4020.6892&2006 Oct. 12&$3 \times 2700$&197&185\\
Fornax&4.3&4021.6101&2006 Oct. 13&$3 \times 2000$&224&218\\
&&&&$1 \times 1700$\\
Fornax&9.2&4022.7825&2006 Oct. 14&$3 \times 2700$&192&109\\
&&&&$1 \times 1800$\\
Fornax&54&4023.6057&2006 Oct. 15&$2 \times 2700$&80&74\\
&&&&$1 \times 2000$\\
Fornax&54&4024.5654&2006 Oct. 16&$1 \times 2700$&80&&repeat from 2006 Oct. 15\\
Fornax&81&4024.6206&2006 Oct. 16&$3 \times 2700$&101&95\\
NGC 7099&1&4025.5010&2006 Oct. 16&$3 \times 300$&59&50&\\
NGC 288&1&4025.5530&2006 Oct. 17&$3 \times 300$&64&60&\\
Fornax&9.3&4025.6353&2006 Oct. 17&$3 \times 2700$&224&205\\
Sculptor&743&4026.5581&2006 Oct. 18&$3 \times 2700$&25&9\\
Fornax&33&4026.6887&2006 Oct. 18&$3 \times 2700$&105&88\\
Fornax&31&4027.5469&2006 Oct. 19&$3 \times 2700$&129&119\\
Fornax&4.4&4027.6787&2006 Oct. 19&$3 \times 2700$&223&216\\
Fornax&70&4027.8198&2006 Oct. 19&$3 \times 1900$&72&60\\
  \enddata
  \label{tab:observations}
\end{deluxetable*}

\section{Reduction of MMFS Spectra}
\label{sec:reduction}

\subsection{Initial Processing}

We extract MMFS spectra and measure radial velocities using standard packages and subroutines available in the IRAF\footnote{IRAF is distributed by the National Optical Astronomy Observatories, which is operated by the Association of Universities for Research in Astronomy, Inc., under cooperative agreement with the National Science Foundation.} astronomical software package.  After $3 \times 3$ binning of the $2$k $\times 4$k detectors at readout, raw data frames from either detector are $683 \times 1365$ pix$^{2}$, and the circular resolution element has FWHM $\sim 3.5$ binned pixels in the spectral direction.  We reduce blue and red spectra independently and following identical procedures.  

After applying overscan and bias corrections, we use the IMCOMBINE task to average the series of target frames for each field.  Individual frames are scaled and weighted to account for differences in exposure time and airmass.  Bad pixels, including most of those affected by cosmic rays, are rejected using a $4 \sigma$ clip around the median of the scaled frames.  We then use the APSCATTER task to subtract scattered light from target frames.  We identify spectral apertures in the well-exposed quartz frame (averaged in the same manner as the target frames) associated with each target field, and then use the APALL task to extract one-dimensional spectra assuming the target, quartz, and Th-Ar frames associated with a given field have identical spectral apertures.
   
\subsection{Wavelength Calibration}
\label{subsec:wavelength}

We determine the wavelength/pixel relation using absoprtion line features in the solar spectrum observed at evening twilight.  In the observed region the solar spectrum contains twice as many useful lines as the Th-Ar spectrum.  The MMFS wavelength/pixel relation is sufficiently stable over an observing run that the benefit of using more lines outweighs disadvantages that arise because twilight exposures are available only at dusk and dawn.  To quantify any drift in the wavelength/pixel relation during the night we measure the relative redshifts of the twilight-calibrated Th-Ar spectra taken with each target field, and we compensate for these redshifts by applying low-order corrections to individual wavelength solutions.  Because this method is a departure from the conventional Th-Ar calibration, we describe it in detail.

\begin{figure}
  \epsscale{1.2}
  \plotone{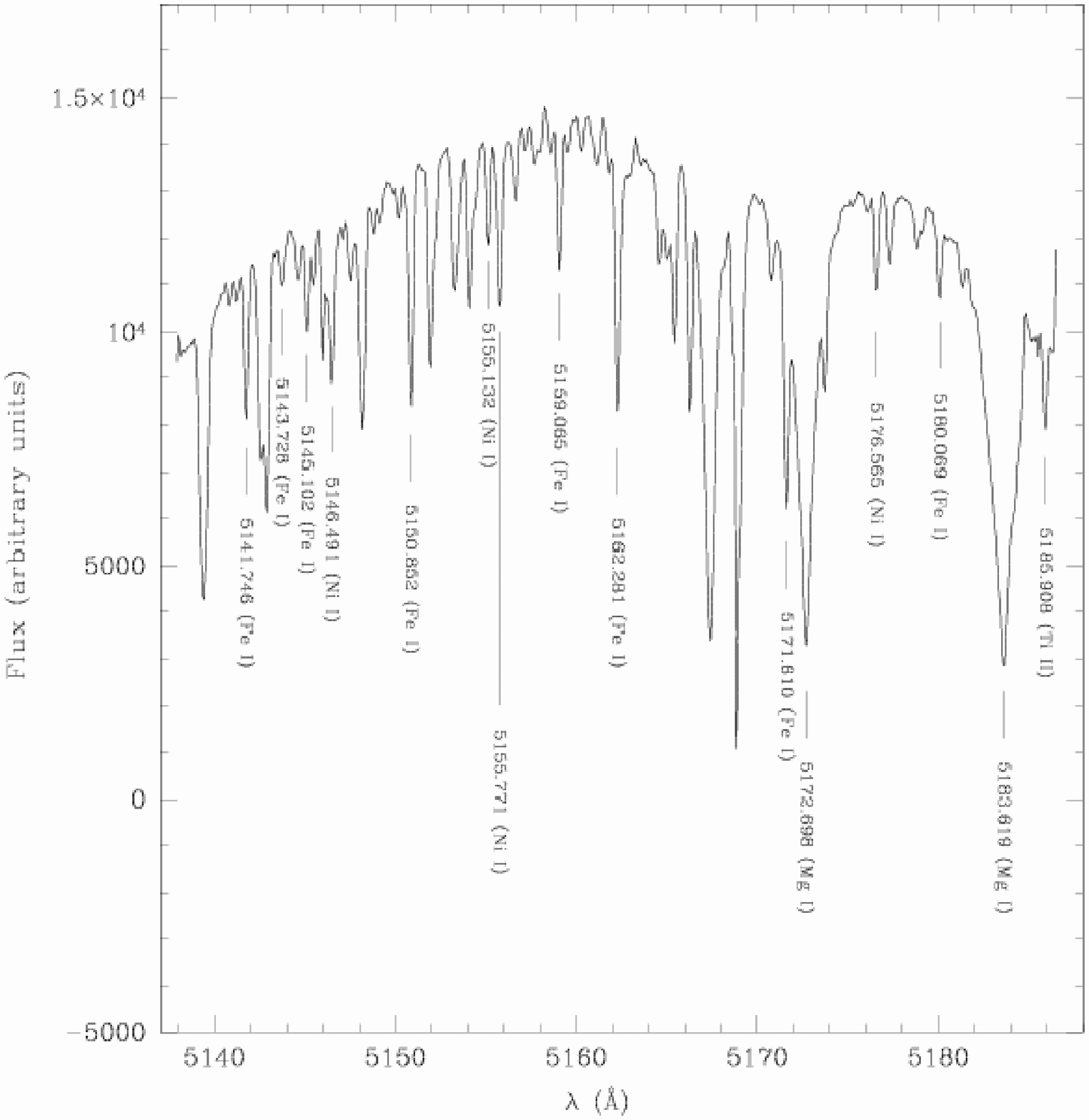}
  \caption{ Example twilight solar spectrum used in wavelength calibration.  Absorption lines used in the wavelength solution are identified.}  
  \label{fig:calibration}
\end{figure}

In our twilight spectra we identify solar absorption lines using the NOAO FTS solar atlas \citep{kurucz84}.  Figure \ref{fig:calibration} displays an example twilight-solar spectrum and labels lines used for wavelength calibration.  We exclude several prominent lines (including the MgI line at 5167.327 \AA) because they are blended with nearby lines.  The wavelength solution is determined from a 4$^{th}$-order polynomial fit; residuals have mean rms $\sim 0.01$\AA\ ($0.6$ km s$^{-1}$) for blue spectra and $\sim 0.02$\AA\ ($1.2$ km s$^{-1}$) for red.  For reference, typical residuals from fits to 9 emission lines in our Th-Ar spectra are larger by a factor of two.

\begin{figure*}
  \plotone{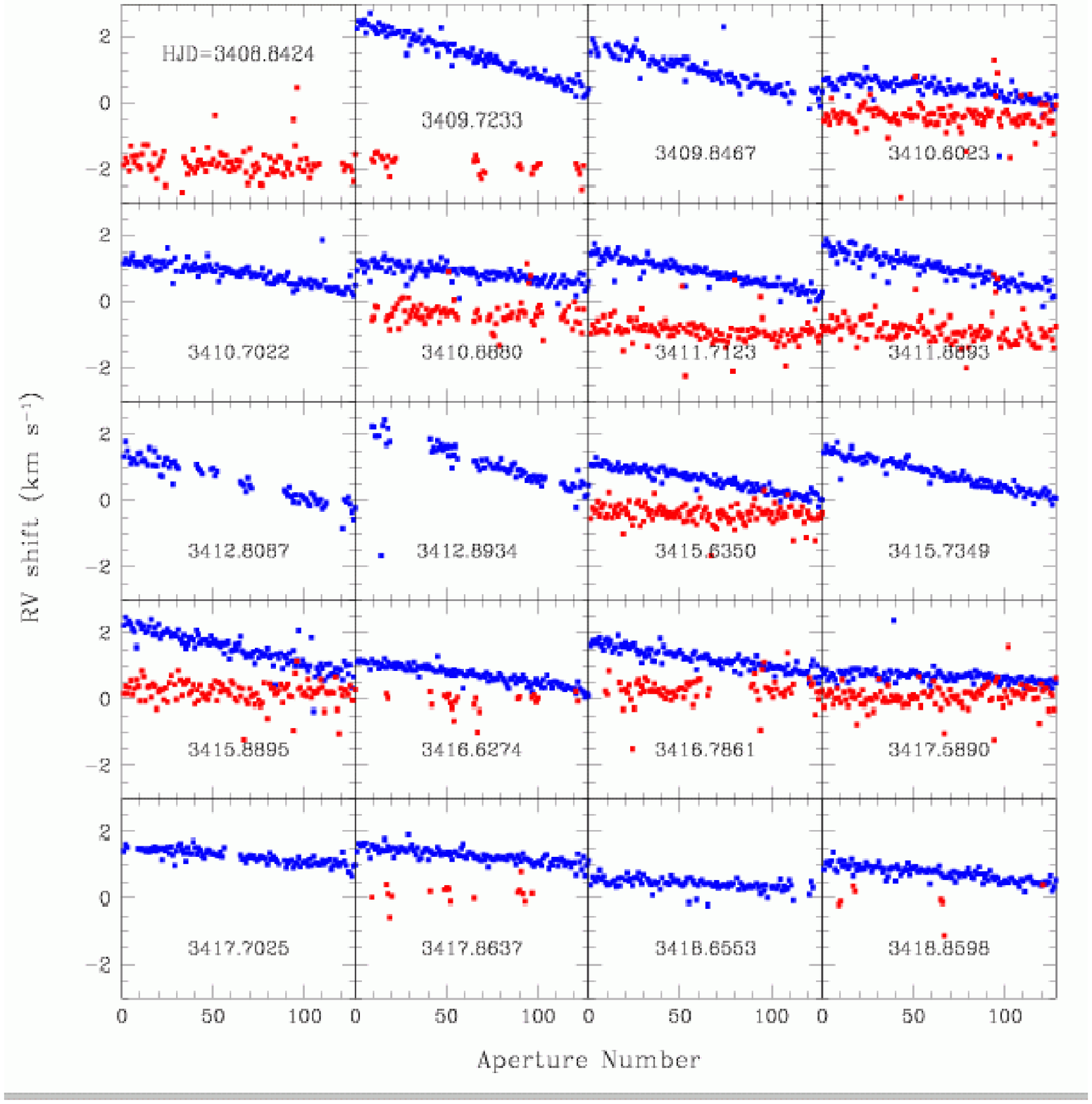}
  \caption{ Stability of the MMFS wavelength/pixel relation.  Plotted for each multispectrum Th-Ar frame in the 2005 February MMFS run is the velocity shift measured via cross-correlation against the Th-Ar frame associated with a single twilight frame.  Prior to cross-correlation, wavelengths in all Th-Ar frames were given identical wavelength/pixel solutions based on solar absorption lines identified in the twilight spectra.  Blue/red points indicate blue/red MIKE detectors.  Each panel indicates the heliocentric Julian date ($-2.45 \times 10^6$ days) at the midpoint of the Th-Ar exposure.}
  \label{fig:tharshift}
\end{figure*}

Next we apply initial wavelength solutions to all frames using the DISPCOR task.  We assign all target, quartz, and Th-Ar frames the (aperture-dependent) wavelength solutions determined from the (temporally) nearest twilight spectrum.  We then use FXCOR to measure zero-point drift in the wavelength/pixel relation (during the night and/or over the course of the run) from shifts in the wavelengths of emission lines in the Th-Ar spectra.  The measured zero-point shifts for Th-Ar frames taken during the observing run in 2005 February are shown as a function of aperture number in Figure \ref{fig:tharshift}.  We find that drift in the wavelength solution results in small ($\leq 2$ km s$^{-1}$) zero-point offsets that depend linearly on aperture number.  We note that the zero-point offsets, and hence the wavelength/pixel relation, remain stable over the span of several nights.  We correct for zero-point drift using the DOPCOR task to shift the wavelength solution in each aperture of the associated target and quartz frames by a value determined from the best linear fit to the zero-point offset vs.\ aperture data.  

\begin{figure}
  \epsscale{1.2}
  \plotone{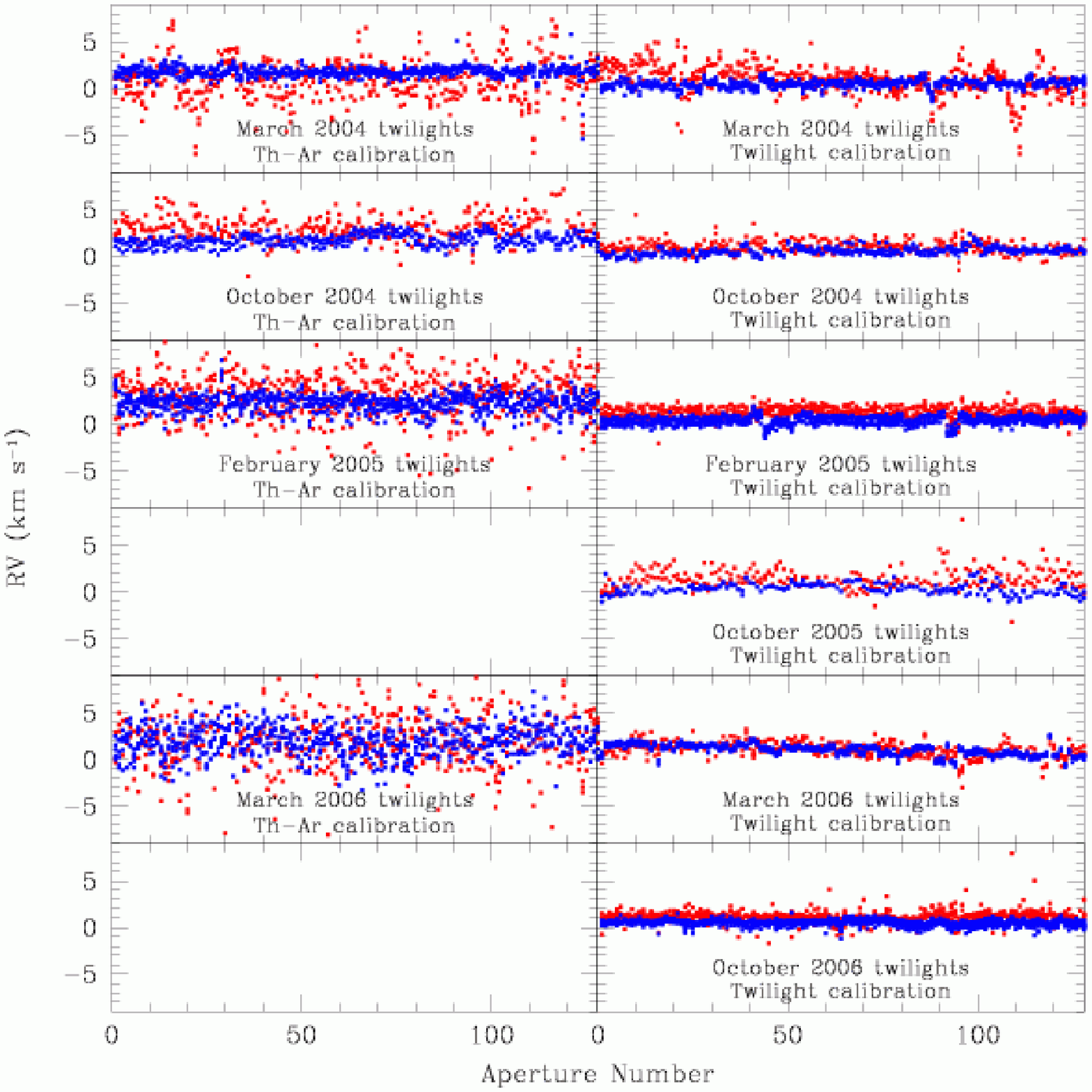}
  \caption{ Comparison of the Th-Ar wavlength calibration method (left panels) with the twilight calibration method (right panels).  For each of the MMFS observing runs, blue/red markers indicate (solar rest frame) radial velocities measured from twilight spectra in each aperture in blue/red channels.  In 2005 October, Th-Ar exposures were not taken immediately after twilight exposures, precluding Th-Ar calibration of those twilight spectra.  Due to the previous comparisons, we did not perform Th-Ar calibration of the 2006 October spectra.}
  \label{fig:twilights}
\end{figure}

Figure \ref{fig:twilights} displays solar radial velocity results measured from the several twilight exposures obtained during each MMFS observing run.  Panels on the right side give velocity results for twilight-calibrated twilight frames.  For comparison, panels on the left side give velocities measured using conventional Th-Ar-calibration.  We note the consistent velocity precision obtained with twilight calibration.  We suspect the degradation in the precision from Th-Ar calibration is related to the fact that the Th-Ar lamp system used exclusively by MMFS has operated at different voltages and with different power supplies over the six observing runs.   The sample standard deviation of velocities measured using twilight calibration has remained steady at $\sim 0.8$ km s$^{-1}$ (blue) and $\sim 1.6$ km s$^{-1}$ (red).  Figure \ref{fig:twilights} also provides evidence of small channel- and in some cases aperture-dependent zero-point offsets.  The red channel tends to yield slightly more positive velocities than the blue channel.  However, it is straightforward to measure and correct for these offsets (see Section \ref{subsec:zeropoint}), which are smaller than the typical velocity measurement error.  

\subsection{Throughput Corrections and Sky Subtraction}
\label{subsec:corrections}
Because fibers differ in throughput by up to a factor of two, we create a response frame by dividing each aperture of the quartz frame by the average (over all apertures) quartz continuum.  Dividing the target frame by the response frame corrects for variations in fiber throughput and pixel sensitivity.  

At this point we use the CONTINUUM task to remove from the target frames any remaining effects due to cosmic rays and/or bad pixels and columns.  We fit a $10^{th}$-order legendre polynomial to the corrected spectrum in each aperture and replace with the function value any pixel values that deviate from the fit by more than $6 \sigma$.  

Sky noise near the Mg-triplet is contributed mainly by scattered solar light.  Left uncorrected, sky contamination produces a secondary velocity signal at $0$ km s$^{-1}$.  During observation of every dSph field we assign $\sim 32$ science fibers ($\sim 16$ per MIKE channel) to regions of blank sky.  We use the SCOMBINE task to average the apertures corresponding to these sky spectra in the target frames.  Prior to averaging we discard the three highest and two lowest values among the sky spectra at each pixel.  Removal of the high pixel values reduces the contribution from resolved background galaxies \citep{wyse92}.  We subtract the average sky spectrum from each aperture in the target frame. 

\begin{figure*}
  \epsscale{1}
  \plotone{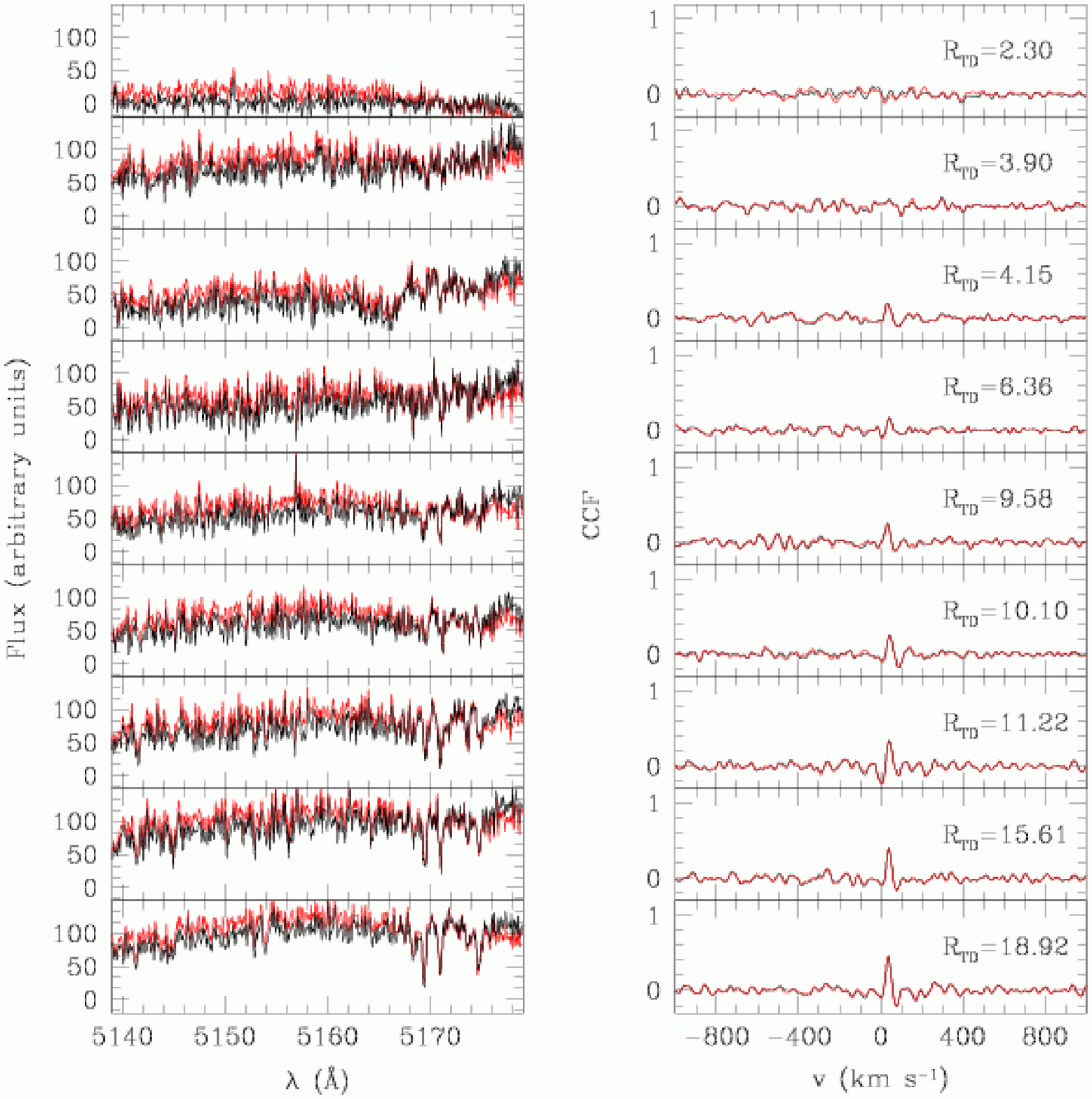}
  \caption{ Processed spectra and corresponding radial velocity cross-correlation functions from observations during a dark night.  Solid black lines in the left-hand panels display examples of fully processed spectra from a Sculptor target field.  Red lines represent each spectrum prior to sky subtraction.  Right-hand panels show each spectrum's cross-correlation against our radial velocity template and list the Tonry-Davis R value measured from the CCF.}
  \label{fig:bscl5spectra}
\end{figure*}

\begin{figure*}
  \epsscale{1}
  \plotone{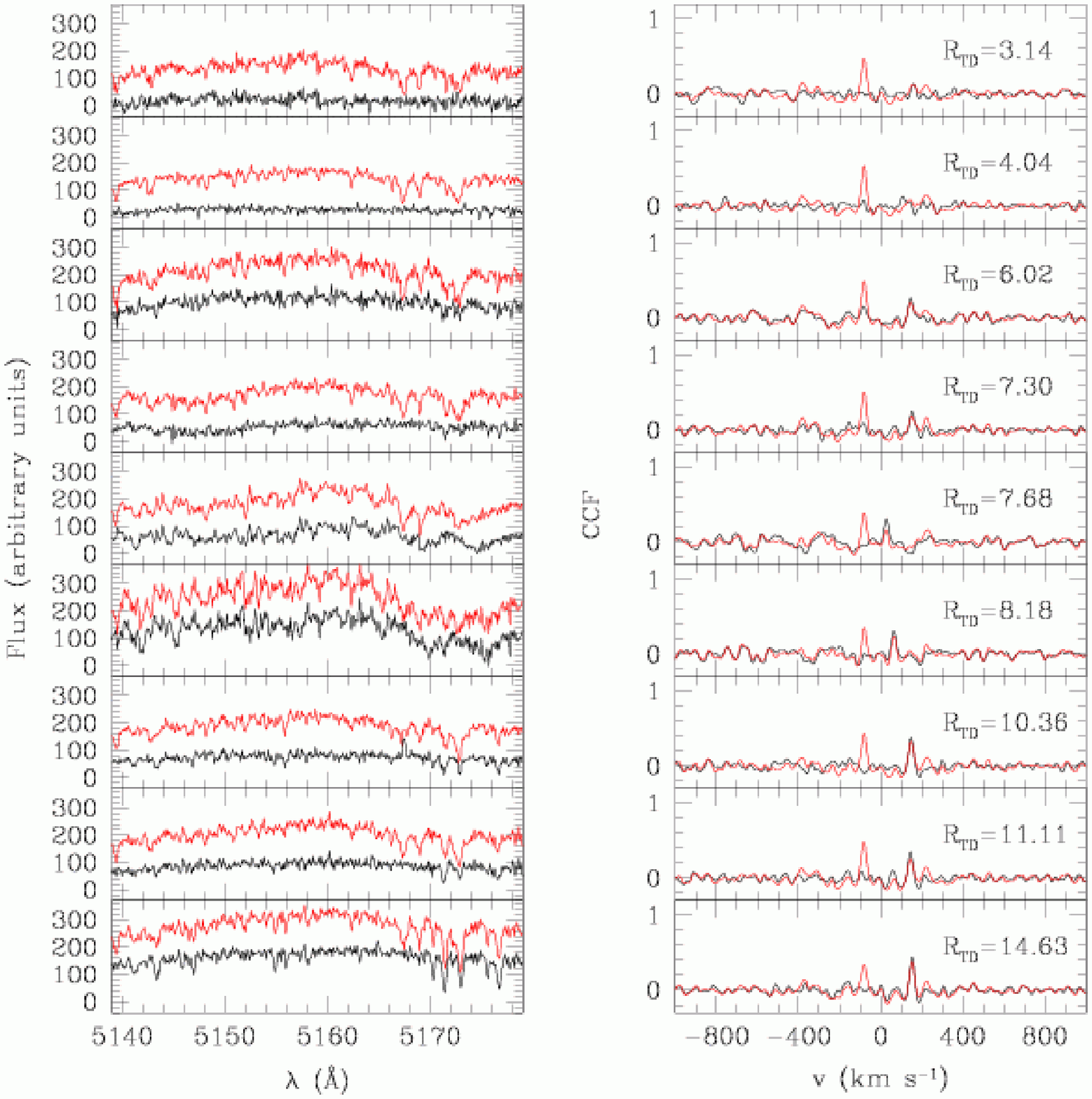}
  \caption{ Panels plot the same relations as in Figure \ref{fig:bscl5spectra}, but for example spectra from a Sextans field observed in the presence of significant moonlight.}
  \label{fig:bsex6spectra}
\end{figure*}

Black, solid lines in the left panels of Figures \ref{fig:bscl5spectra} and \ref{fig:bsex6spectra} display sky-subtracted spectra representing the range of S/N achieved during typical observations of dSph fields.  Red, dotted lines in the left panels of either figure indicate the spectra prior to sky subtraction.  The spectra in Figure \ref{fig:bscl5spectra} were obtained at new moon, while the spectra in Figure \ref{fig:bsex6spectra} were obtained at first-quarter moon.  Prior to sky subtraction, spectra from the latter observation clearly are contaminated by solar absorption lines.  After sky subtraction, residual solar lines typically deviate from the continuum level by an amount similar to the intrinsic continuum noise level.  

\subsection{Final Processing}
\label{subsec:finalprocessing}

In most MMFS observing runs there have been one or more fields for which we deemed the original observation useful but insufficient (see Table 1).  Reasons range from poor observing conditions to inadequate exposure time available prior to morning twilight.  In such cases we re-observed the field using fiber assignments identical to those of the original observation.  During the reduction steps already described, repeat observations---often separated by several days from the original observation---are treated independently to allow for drift in the wavelength solution and different sky levels.  At this point we combine all observations of the same field taken during the same run.  First we use the DOPCOR task to correct the wavelength solutions of repeat observations for the changing component of Earth's motion (typical shifts are less than 1 km s$^{-1}$) toward the target field.  The SCOMBINE task then averages spectra in corresponding apertures of the original and repeat target frames.  Spectra in individual apertures are weighted by the mean pixel value prior to computing the aperture average.  

In the final processing step prior to velocity measurement, we remove the continuum shape of the spectra.  The CONTINUUM task fits a $10^{th}$-order legendre polynomial to the spectrum in each aperture of the target frames, then subtracts the function values from pixel values to produce the final spectra.

\subsection{Signal-to-Noise Ratio}
\label{snratio}

The variance in our spectra has independent components contributed by source, sky background, and read noise.  For the $i^{th}$ pixel, let $N_{S,i}$ represent the number of detected photons from the target star, let $N_{B,i}$ represent the number of detected photons from the sky background, and let $\sigma_R^2$ be the noise associated with detector readout (typically between 2-4 electrons per pixel for MIKE).  In practice we estimate $N_{B,i}$ from $s$ individual sky spectra.  Assuming the estimate of $N_{B,i}$ contributes a variance component $N_{B,i}/s$, the signal-to-noise ratio (S/N) at the $i^{th}$ pixel is given by 
\begin{equation} 
\biggl [\frac{S}{N} \biggr]_i=\frac{N_{S,i}}{\sqrt{N_{S,i}+N_{B,i}+N_{B,i}/s+\sigma_R^2}}.
\label{eq:snratio}
\end{equation}

\begin{figure}
  \epsscale{1.2}
  \plotone{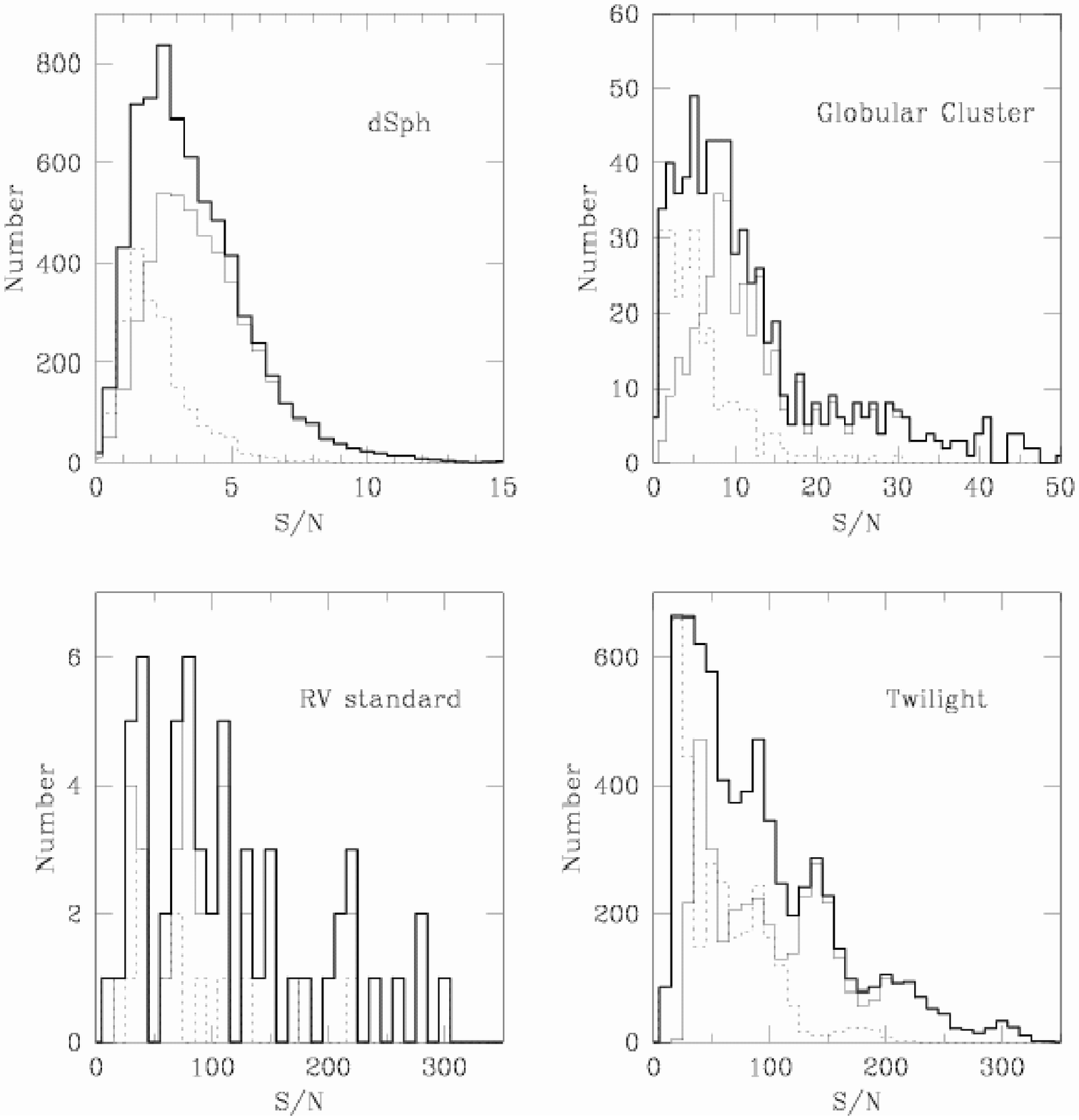}
  \caption{ Distributions of the mean signal-to-noise ratio (S/N) per pixel, calculated over the wavelength range $5141-5177$ \AA.  Distributions are plotted separately for dSph, globular cluster, RV standard and solar twilight spectra.  In each panel the thin solid (dotted) lines give distributions for spectra obtained with MIKE's blue (red) channels, while the thick solid line is the combined distribution.  The resolution element is 3.5 pixels and the dispersion is 0.08 \AA /pix (0.12 \AA /pix) for blue (red) channels.}  
  \label{fig:snratio}
\end{figure}

We estimate source and background photon fluxes from smooth fits to the the continua of our sky-subtracted and averaged sky spectra, respectively.  At the $i^{th}$ pixel the continuum fit to the sky-subtracted spectrum has digital unit (DU) value\footnote{We follow the convention whereby $\hat{q}$ denotes the estimate of $q$.} $n_{S,i}=\hat{N}_{S,i}/(R_iG)$, where $R_i$ is the value of the response frame (Section \ref{subsec:corrections}) and $G$ is the detector gain (between 0.5 and 1.5 electrons/DU over the several MMFS observing runs).  The continuum fit to the averaged sky spectrum has DU value $n_{B,i}=\hat{N}_{B,i}/(R_iG)$.  From the estimates $\hat{N}_{S,i}$ and $\hat{N}_{B,i}$ we calculate S/N at each pixel using Equation \ref{eq:snratio}.  Figure \ref{fig:snratio} plots, for all spectra that eventually yield an acceptable velocity measurement, the distribution of the mean S/N calculated from pixels spanning the range $5141 - 5177$ \AA.  The median S/N per pixel among dSph spectra acquired with MIKE's blue (red) channel is 3.7 (2.0) per pixel.  Multiplying these values by $\sqrt{3.5}$ gives the mean S/N per resolution element.  

\subsection{Measurement of Radial Velocities}
\label{subsec:ccf}

We measure radial velocity using the FXCOR task, which cross-correlates every continuum-subtracted spectrum against a stellar template spectrum of known redshift.  For the template we use the sum of 428 spectra from bright, late-type stars, including 155 spectra of radial velocity standard stars and 273 spectra of stars from 7 Milky Way globular clusters.  All spectra contributing to the template were obtained with echelle spectrographs, either during a previous radial velocity study \citep{walker06a} or with MMFS as part of the present study, and all have spectral resolution equal to or higher than those of the target spectra.  We co-add these individual spectra after shifting each to a common redshift.  
  
For cross-correlation, all spectra are re-binned linearly with $\log$($\lambda/$\AA) at a common dispersion of $3.11 \times 10^{-6}$ pix$^{-1}$.  Fourier transforms of the target and template spectra are filtered such that low-frequency continuum residuals and undersampled features at high frequencies do not contribute to the resulting cross-correlation function (CCF).  The right panels of Figures \ref{fig:bscl5spectra} and \ref{fig:bsex6spectra} display CCFs calculated from the spectra shown in the left panels.  The relative velocity between object and template is the velocity value at the center of a Gaussian profile fit to the tallest CCF peak (within a window of width 700 km s$^{-1}$, centered on the velocity of the target dSph).  From the CCF, the known redshift of the template, the time of observation, and coordinates of the object, FXCOR returns the radial velocity in the solar rest frame.  

\subsection{Velocity Zero Point}
\label{subsec:zeropoint}

The solar velocity measurements in Figure \ref{fig:twilights} show evidence of small residual offsets that depend on channel and aperture.  We correct for these offsets by demanding that the many repeat velocity measurements obtained from solar twilight spectra yield mean velocities of 0 km s$^{-1}$ in each aperture in each run.  We obtained sets of $2-7$ independent velocity measurements per aperture, per observing run, from solar twilight spectra (for a total of 6816 twilight spectra).  We subtract the mean velocity of each such set from all velocity measurements (including those from twilight exposures, science targets and radial velocity standard stars) obtained using the same aperture in the same observing run; this places velocities measured in a given run on a common zero point.  The mean value subtracted from blue (red) apertures is $0.48$ km s$^{-1}$ ($1.07$ km s$^{-1}$ ), and the corrections have a standard deviation of $0.48$ km s$^{-1}$ ($0.78$ km s$^{-1}$).  The standard deviation from zero among corrected twilight RVs is 0.26 km s$^{-1}$ (0.60 km s$^{-1}$) for blue (red) apertures; we adopt this value as the baseline velocity error in the error analysis (Section \ref{subsec:errors}).

\begin{figure}
  \epsscale{1.2}
  \plotone{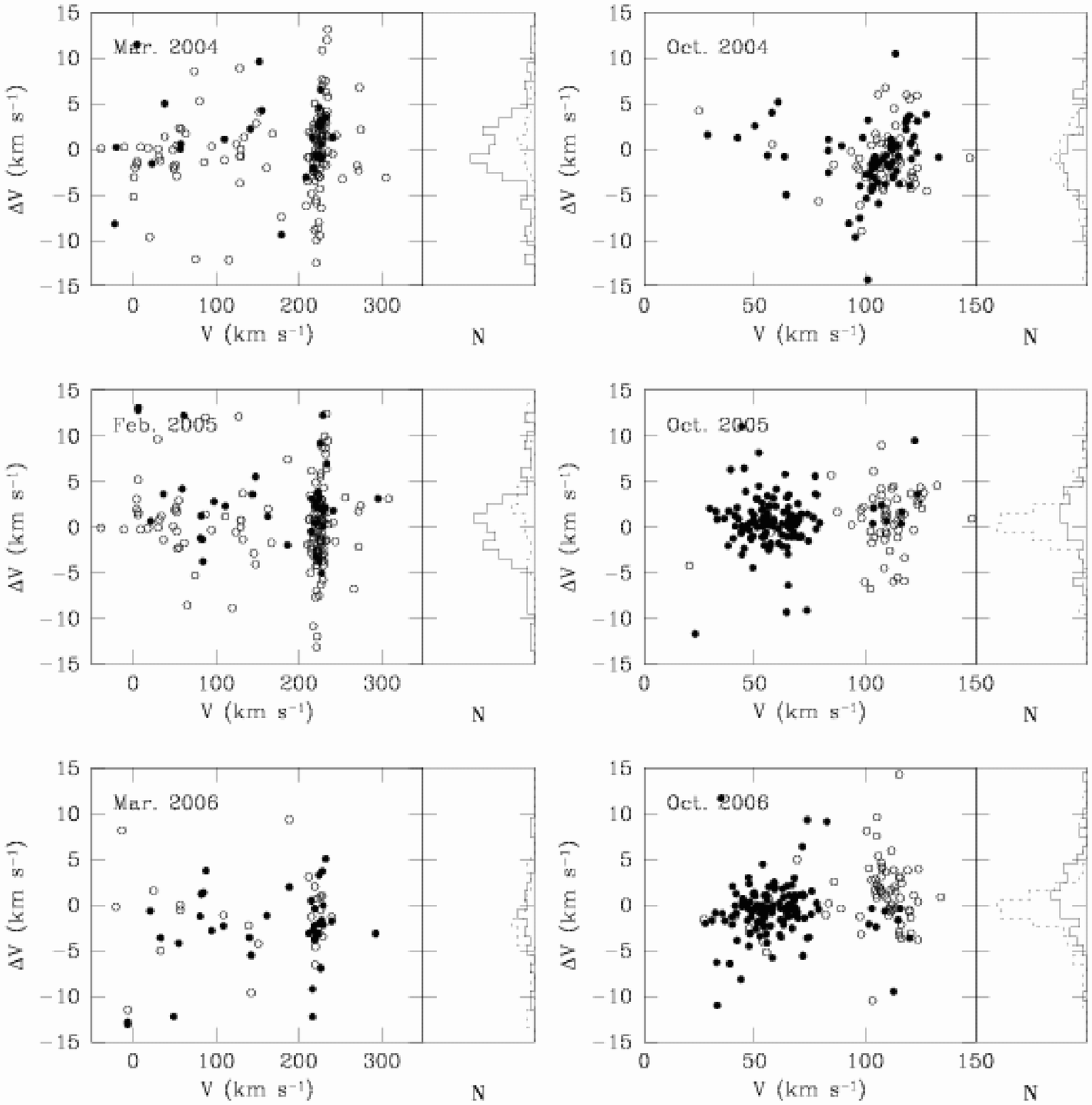}
  \caption{ Distribution of measurement deviations for stars measured with the blue channel in multiple observing runs.  Left-hand sub-panels plot velocity deviation as a function of the (chronologically) first measurement; histograms in the right-hand subpanels give the distribution of $\Delta$V.   We observed a given star in up to three different runs.  Filled circles and solid histograms result from comparison with the earlier of the two other runs; open circles and dotted histograms result from comparison with the latter of the two other runs.}
  \label{fig:runshiftb}
\end{figure}

\begin{figure}
  \epsscale{1.2}
  \plotone{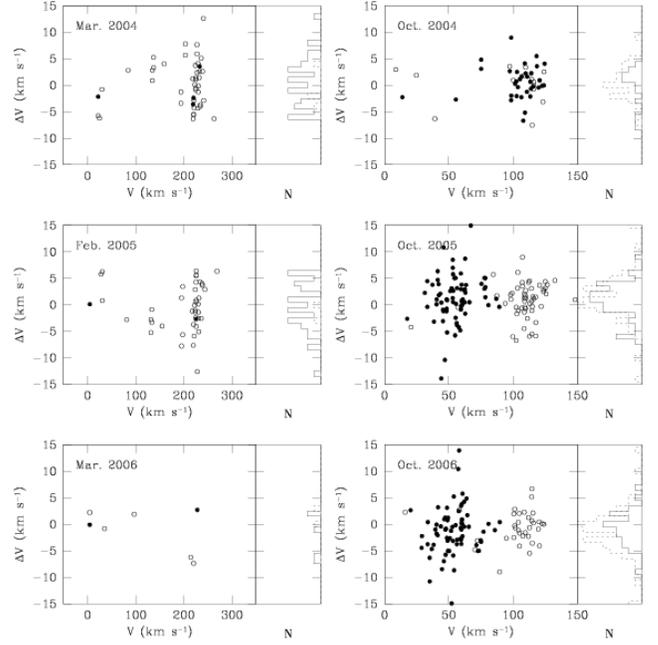}
  \caption{ Same as Figure \ref{fig:runshiftb} but for stars observed with the red channel in multiple runs.}
  \label{fig:runshiftr}
\end{figure}

Because the template spectrum is composed of late-type stellar spectra, velocities measured from solar spectra are susceptible to object/template mismatch.  While we consider the the zero-point corrections described above to be sufficient for removing channel-and aperture-dependent offsets within a given run, we next use dSph targets observed with the same channel in two or more runs to correct for variation in the velocity zero point between observing runs.  For such stars measured on blue (red) channels, Figure \ref{fig:runshiftb} (Figure \ref{fig:runshiftr}) plots the difference between measured velocities as a function of the earlier measured velocity.  Histograms indicate the distribution of measurement differences.  In order to bring the center of each distribution to zero we apply constant offsets to all velocities measured in a given channel during a given run.  Arbitrarily we choose the run/channel having the smallest mean velocities to serve as the absolute zero point.  Table \ref{tab:runshift} summarizes the applied corrections.  For each MMFS run listed in column 1, columns 2 and 3 list the number of stars on blue and red channels that were observed in the same channel during at least one other run.  Columns 3 and 4 list the blue and red offsets added to every velocity from the run.  The maximum correction is $1.7$ km s$^{-1}$, slightly smaller than the median measurement error (Section \ref{subsec:errors}).


\renewcommand{\arraystretch}{0.6}
\begin{deluxetable}{cccccc}
  \tabletypesize{\scriptsize}

  \tablewidth{0pc}
  \tablecaption{ Run-dependent velocity zero-point corrections}
  \tablehead{\colhead{MMFS run}&\colhead{N$_{blue}$}&\colhead{N$_{red}$}&\colhead{Blue correction}&\colhead{Red correction}\\
    \colhead{}&\colhead{}&\colhead{}&\colhead{(km s$^{-1}$)}&\colhead{(km s$^{-1}$)}
}
  \startdata
  Mar. 2004&178&45&-1.65&-1.45\\
  Oct. 2004&125&37&0.0&-0.57\\
  Feb. 2005&194&44&-1.59&-1.31\\
  Oct. 2005&185&73&-1.70&0.0\\
  Mar. 2006&68&2&0.0&0.0\\
  Oct. 2006&198&110&-1.08&0.0\\
  \enddata
  \label{tab:runshift}
\end{deluxetable}     

As a check on the absolute RV zero point we consider 58 velocity measurements of 22 RV standard stars (after applying the zero point shifts described above).  Table \ref{tab:std_veldata1} lists the individual MMFS velocity measurements for each standard, and for comparison lists previously published velocity measurements, from \citet{edo91,udry99,beers00,walker06a}.  The mean deviation between the measured RV and the published value is 0.07 km s$^{-1}$, which indicates the zero point of the MMFS velocities is consistent with RV standards.  In the case of HD 45282, \citet{beers00} estimate a measurement error of $\pm 10$ km s$^{-1}$; thus despite a deviation of 6 km s$^{-1}$, the MMFS measurement is consistent with the published value.  Repeat MMFS measurements of RV standards show excellent agreement; the largest deviation among repeat measurements is 3.4 km s$^{-1}$, and 93\% of repeat measurements differ by less than 3 km s$^{-1}$.    

\renewcommand{\arraystretch}{0.6}
\begin{deluxetable*}{lcrrrrrrrr}
  \tabletypesize{\scriptsize}
  \tablewidth{0pc}
  \tablecaption{ MMFS Velocities of Radial Velocity Standards}
 \label{tab:std_veldata1}
 \tablehead{\colhead{ID}&\colhead{HJD}&\colhead{$V_{published}$}&\colhead{$V$}&\colhead{$V-V_{published}$}\\
   \colhead{}&\colhead{(-2.45 $\times 10^6$ days)}&\colhead{(km s$^{-1}$)}&\colhead{(km s$^{-1}$)}&\colhead{(km s$^{-1}$)}
}
\startdata
HD 83212  &   3088.6241&$  110.0\pm    1.0$\tablenotemark{3}&$  107.4\pm    0.4$&$   -2.6\pm    1.1$\\
&  3088.6276&$  110.0\pm    1.0$&$  107.5\pm    0.4$&$   -2.5\pm    1.1$\\
&  3088.6321&$  110.0\pm    1.0$&$  109.2\pm    0.7$&$   -0.8\pm    1.2$\\
&  3088.6363&$  110.0\pm    1.0$&$  109.3\pm    0.7$&$   -0.7\pm    1.2$\\
HD 103545 &   3088.6441&$  177.0\pm    1.5$\tablenotemark{4}&$  178.9\pm    3.3$&$    1.9\pm    3.6$\\
&  3088.6491&$  177.0\pm    1.5$&$  176.9\pm    1.3$&$   -0.1\pm    2.0$\\
&  3090.6633&$  177.0\pm    1.5$&$  177.6\pm    1.2$&$    0.6\pm    1.9$\\
&  3090.6666&$  177.0\pm    1.5$&$  179.5\pm    3.4$&$    2.5\pm    3.7$\\
&  3409.8842&$  177.0\pm    1.5$&$  177.2\pm    1.5$&$    0.2\pm    2.1$\\
&  3417.8769&$  177.0\pm    1.5$&$  176.9\pm    1.2$&$   -0.1\pm    1.9$\\
&  3418.8698&$  177.0\pm    1.5$&$  177.8\pm    1.4$&$    0.8\pm    2.0$\\
HD 45282 &    3089.4900&$  301.0\pm   10.0$\tablenotemark{3}&$  307.1\pm    2.4$&$    6.1\pm   10.3$\\
HD 80170  &   3089.6553&$    0.5\pm    0.2$\tablenotemark{2}&$    0.5\pm    0.7$&$    0.0\pm    0.8$\\
&  3089.6584&$    0.5\pm    0.2$&$   -1.6\pm    0.3$&$   -2.1\pm    0.4$\\
&  3092.6314&$    0.5\pm    0.2$&$   -2.8\pm    0.3$&$   -3.3\pm    0.4$\\
&  3092.6335&$    0.5\pm    0.2$&$   -0.1\pm    0.7$&$   -0.6\pm    0.8$\\
HD 92588   &  3089.6641&$   42.5\pm    0.3$\tablenotemark{2}&$   40.8\pm    0.4$&$   -1.8\pm    0.5$\\
&  3089.6683&$   42.5\pm    0.3$&$   42.1\pm    0.7$&$   -0.4\pm    0.8$\\
&  3091.6315&$   42.5\pm    0.3$&$   41.9\pm    0.7$&$   -0.6\pm    0.8$\\
&  3091.6343&$   42.5\pm    0.3$&$   41.1\pm    0.3$&$   -1.4\pm    0.4$\\
HD 124358  &  3090.8064&$  325.0\pm    1.0$\tablenotemark{3}&$  323.6\pm    2.7$&$   -1.4\pm    2.9$\\
&  3090.8111&$  325.0\pm    1.0$&$  323.0\pm    0.7$&$   -2.0\pm    1.2$\\
HD 157457  &  3090.8115&$   17.8\pm    0.3$\tablenotemark{2}&$   16.1\pm    0.3$&$   -1.7\pm    0.4$\\
&  3090.8134&$   17.8\pm    0.3$&$   16.6\pm    0.8$&$   -1.2\pm    0.8$\\
HD 118055  &  3091.6431&$ -101.0\pm    1.0$\tablenotemark{3}&$ -101.8\pm    0.4$&$   -0.8\pm    1.1$\\
&  3091.6471&$ -101.0\pm    1.0$&$ -102.2\pm    0.9$&$   -1.2\pm    1.3$\\
&  3415.9002&$ -101.0\pm    1.0$&$ -103.2\pm    0.4$&$   -2.2\pm    1.1$\\
&  3417.8831&$ -101.0\pm    1.0$&$ -102.7\pm    0.4$&$   -1.7\pm    1.1$\\
&  3801.8875&$ -101.0\pm    1.0$&$ -100.2\pm    0.4$&$    0.8\pm    1.1$\\
&  3801.8891&$ -101.0\pm    1.0$&$ -100.1\pm    0.4$&$    0.9\pm    1.1$\\
HD 176047  &  3287.5020&$  -42.5\pm    0.2$\tablenotemark{2}&$  -41.8\pm    0.3$&$    0.7\pm    0.4$\\
HD 196983  &  3287.5157&$   -9.1\pm    0.3$\tablenotemark{2}&$   -8.3\pm    0.3$&$    0.8\pm    0.4$\\
&  4018.5032&$   -9.1\pm    0.3$&$   -9.3\pm    0.3$&$   -0.2\pm    0.4$\\
&  4027.4955&$   -9.1\pm    0.3$&$   -9.8\pm    0.3$&$   -0.7\pm    0.4$\\
&  4027.5121&$   -9.1\pm    0.3$&$   -7.0\pm    0.7$&$    2.1\pm    0.7$\\
HD 219509  &  3287.5246&$   67.5\pm    0.5$\tablenotemark{2}&$   66.3\pm    1.0$&$   -1.2\pm    1.1$\\
CPD-432527 &  3287.8984&$   19.7\pm    0.9$\tablenotemark{2}&$   20.4\pm    0.3$&$    0.7\pm    0.9$\\
&  3410.5182&$   19.7\pm    0.9$&$   17.9\pm    0.3$&$   -1.8\pm    0.9$\\
&  3665.8825&$   19.7\pm    0.9$&$   19.3\pm    0.3$&$   -0.4\pm    0.9$\\
HD 48381  &   3288.8948&$   40.5\pm    0.2$\tablenotemark{2}&$   41.6\pm    0.3$&$    1.1\pm    0.3$\\
&  3289.8915&$   40.5\pm    0.2$&$   43.2\pm    0.3$&$    2.7\pm    0.3$\\
&  3289.8939&$   40.5\pm    0.2$&$   43.4\pm    0.3$&$    2.9\pm    0.3$\\
&  3666.8791&$   40.5\pm    0.2$&$   39.9\pm    0.3$&$   -0.6\pm    0.3$\\
SAO 201636 &  3408.8782&$  264.3\pm    1.4$\tablenotemark{4}&$  264.8\pm    1.0$&$    0.5\pm    1.7$\\
&  3409.8670&$  264.3\pm    1.4$&$  264.2\pm    0.7$&$   -0.1\pm    1.6$\\
HD 83516  &   3409.8590&$   43.5\pm    0.2$\tablenotemark{2}&$   41.8\pm    0.3$&$   -1.7\pm    0.3$\\
&  3412.9035&$   43.5\pm    0.2$&$   42.0\pm    0.3$&$   -1.5\pm    0.3$\\
HD 111417  &  3410.8967&$  -19.1\pm    0.2$\tablenotemark{2}&$  -19.1\pm    0.6$&$    0.0\pm    0.7$\\
&  3803.8877&$  -19.1\pm    0.2$&$  -18.7\pm    0.6$&$    0.4\pm    0.7$\\
HD 43880   &  3411.5240&$   43.6\pm    2.4$\tablenotemark{1}&$   40.6\pm    0.3$&$   -2.9\pm    2.4$\\
&  3412.5257&$   43.6\pm    2.4$&$   40.3\pm    0.3$&$   -3.3\pm    2.4$\\
&  3801.5077&$   43.6\pm    2.4$&$   43.3\pm    0.3$&$   -0.3\pm    2.4$\\
&  3801.5090&$   43.6\pm    2.4$&$   43.2\pm    0.3$&$   -0.4\pm    2.4$\\
HD 93529  &   3411.9006&$  143.0\pm    1.0$\tablenotemark{3}&$  144.3\pm    0.3$&$    1.3\pm    1.1$\\
HD 23214  &   3415.5220&$   -5.1\pm    0.9$\tablenotemark{4}&$   -6.8\pm    0.3$&$   -1.7\pm    1.0$\\
HD 21581  &   3416.5260&$  151.3\pm    1.3$\tablenotemark{4}&$  151.7\pm    0.5$&$    0.4\pm    1.4$\\
SAO 217998 &  3417.5155&$   19.0\pm    1.6$\tablenotemark{4}&$   17.9\pm    0.3$&$   -1.1\pm    1.6$\\
HD 223311  &  4018.5258&$  -20.2\pm    0.3$\tablenotemark{2}&$  -20.0\pm    0.4$&$    0.2\pm    0.5$\\
\enddata
\tablenotetext{1}{\citet{edo91}}
\tablenotetext{2}{\citet{udry99}}
\tablenotetext{3}{\citet{beers00}}
\tablenotetext{4}{\citet{walker06a}}
 \end{deluxetable*}

\subsection{Quality Control}
\label{subsec:qc}

The right panels of Figures \ref{fig:bscl5spectra} and \ref{fig:bsex6spectra} demonstrate the range in quality of the CCF peaks we obtain from the MMFS spectra.  In order to help eliminate poorly measured velocities from the sample, we inspect all CCFs by eye and assign to each a pass/fail grade according to the following criteria.  First, the tallest peak (within the search window centered on the dSph systemic velocity) in a satisfactory CCF must be unambiguous.  CCFs in which the height of one or more secondary peaks reaches 0.8 times that of the tallest peak are flagged as unsatisfactory.  Second, the tallest peak in a satisfactory CCF  must be symmetric about the measured velocity.  Gaussian fits to asymmetric peaks can result in substantial velocity errors; we therefore flag as unsatisfactory those CCFs having significant residuals with respect to the Gaussian fit.  According to these criteria, the CCFs in the top two panels of both Figure \ref{fig:bscl5spectra} and Figure \ref{fig:bsex6spectra} are considered unsatisfactory.  

\begin{figure}
  \epsscale{1.2}
  \plotone{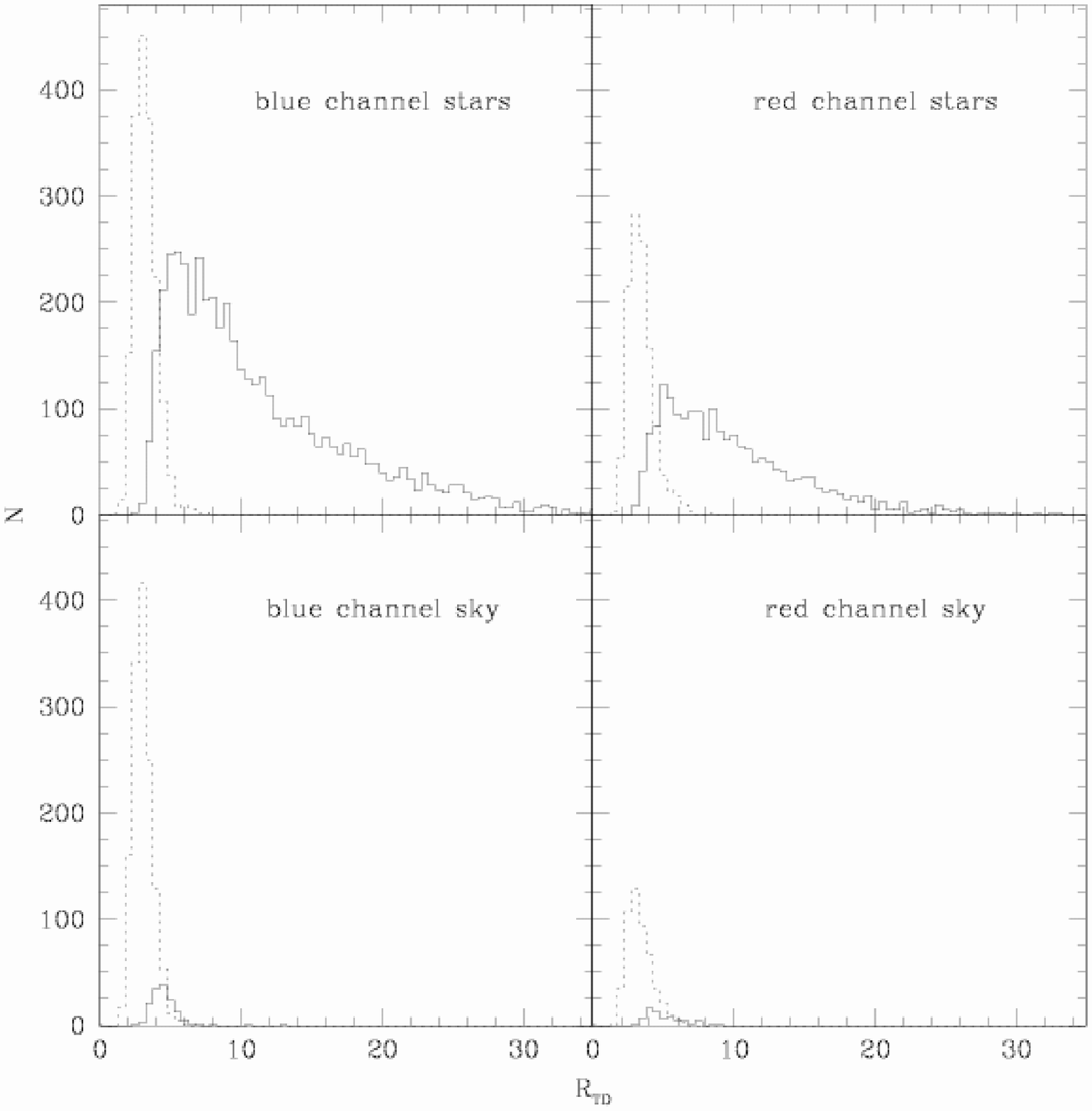}
  \caption{ Distributions of $R_{TD}$ associated with measured cross-correlation functions.  Panels indicate whether the distribution corresponds to stellar or blank-sky spectra obtained in MIKE's blue or red channel.  In each panel, the solid/dotted histogram is the distribution of $R_{TD}$ associated with CCFs judged by eye inspection to be satisfactory/unsatisfactory.}
  \label{fig:tdr}
\end{figure}

A second, quantitative quality control filter is provided by the Tonry-Davis value, $R_{TD}$ (calculated by FXCOR), which indicates the height of the tallest CCF peak relative to the average CCF peak \citep{tonry79}.  Figure \ref{fig:tdr} plots, separately for all the dSph stellar and blank-sky spectra measured in blue and red channels, the distributions of $R_{TD}$ associated with CCFs judged by eye to be satisfactory (solid histograms) and unsatisfactory (dotted histograms).  We expect blank-sky spectra to yield poor CCFs; this is generally the case, as indicated by the small number of sky spectra judged by eye to give satisfactory CCFs and the relatively low $R_{TD}$ values associated with these measurements.  Among stellar spectra, qualitatively satisfactory CCFs have $2.8 \leq R_{TD} \leq 40$, with median $R_{TD}=9.12$.  We define a critical value, $R_{TD,c}$, to be the value of $R_{TD}$ at which, absent other information, the probability that a stellar CCF is satisfactory equals the probability that it is unsatisfactory.  For spectra obtained with MIKE's blue channel this occurs at $R_{TD,c}=4.0$; for the red channel $R_{TD,c}=4.5$.  On the basis of Figure \ref{fig:tdr}, we allow a given stellar velocity measurement into the final sample only if it is derived from a CCF that 1) is judged by eye to be satisfactory, and 2) has $R_{TD}\geq R_{TD,c}$.  

\section{Velocity Measurement Errors}
\label{subsec:errors}

\subsection{Gaussian Errors}
\label{subsec:gaussianerrors}

For the majority of stars in our sample we have obtained only a single velocity measurement and direct estimation of the measurement error is impossible.  Instead we use the many existing cases of independent repeat measurements to relate the errors to quantifiable features of individual spectra.  Let $V_{ij}$ be the $i^{th}$ of $n_j$ independent radial velocity measurements obtained for the $j^{th}$ of $N$ stars in the data set, and suppose the $j^{th}$ star has true radial velocity $V_{*j}$.  We adopt a model in which the velocity measurement error, $V_{ij}-V_{*j}$, is the sum of two independent Gaussian components.  First, noise in the stellar spectrum gives rise to noise in the cross-correlation function.  We assume the error component due to CCF noise follows a Gaussian distribution with variance $\sigma_{CCF}^2$.  The finite spectral resolution of the instrument contributes a second error component that is independent of signal-to-noise ratio and defines the maximum precision of the instrument.  We assume this ``baseline'' error follows a Gaussian distribution with constant variance, $\sigma_0^2$.  The error in a given velocity measurement is then
\begin{equation}
  V_{ij}-V_{*j}=(\sigma_{CCF,ij}^2+\sigma_0^2)^{1/2}\epsilon_{ij}.
  \label{eq:errormodel}
\end{equation}
The values $\epsilon_{ij}$, which account for the inherent randomness of measurement, follow the standard normal (Gaussian with mean zero and unit variance) distribution.  The resulting error bars are symmetric about $V_{ij}$, and given by $\pm \sigma_{V,ij}$, where $\sigma_{V,ij}=(\sigma_{CCF,ij}^2+\sigma_0^2)^{1/2}$.  

We model $\sigma_{CCF}$ as in \citet{walker06a}:
\begin{equation}
  \sigma_{CCF,ij}=\frac{\alpha}{(1+R_{TD,ij})^x}.
  \label{eq:tdr}
\end{equation}
Except for the introduction of the parameter $x$, this is essentially the model of \citet{tonry79}, in which $\sigma_{CCF} \propto (1+R_{TD})^{-1}$.  

After substituting for $\sigma_{CCF}$ in Equation \ref{eq:errormodel}, taking the base-ten logarithm gives
\begin{equation}
  \log[(V_{ij}-V_{*j})^2]=\log \biggl[\frac{\alpha^2}{(1+R_{TD,ij})^{2x}}+\sigma_0^2 \biggr ] + \log[\epsilon_{ij}^2].
  \label{eq:logerrormodel1}
\end{equation}
For normally distributed $\epsilon_{ij}$, the values $\epsilon_{ij}^2$ follow the $\chi^2$ distribution (one degree of freedom).  From Monte Carlo simulations, $\log [\epsilon_{ij}^2]$ has mean value $\langle \log[\epsilon_{ij}^2] \rangle = -0.55$.  If we define $\delta_{ij}\equiv \log [\epsilon_{ij}^2] +0.55$, then 
\begin{equation}
  \log[(V_{ij}-V_{*j})^2]=\log \biggl[\frac{\alpha^2}{(1+R_{TD,ij})^{2x}}+\sigma_0^2 \biggr ] + \delta_{ij}-0.55, 
  \label{eq:logerrormodel2}
\end{equation}
and $\langle \delta_{ij}\rangle = 0$.  For the baseline variance we adopt the standard deviation of velocities measured from twilight spectra.  The CCFs from these spectra generally have $R_{TD} \geq 30$, a regime in which $\sigma_{CCF}$ is negligible.  From Section \ref{subsec:zeropoint}, $\sigma_0=0.26$ km s$^{-1}$ ($\sigma_0=0.60$ km s$^{-1}$) for blue (red) channels.  The parameter pair $\{x,\alpha\}$ can then be estimated, via nonlinear regression, as that which minimizes the sum
\begin{equation}
  \displaystyle\sum_{j=1}^N\displaystyle\sum_{i=1}^{n_j} \biggl (\log[(V_{ij}-V_{*j})^2]- \biggl [ \log \biggl (\frac{\alpha^2}{(1+R_{TD,ij})^{2x}}+\sigma_0^2 \biggr )-0.55 \biggr ] \biggr )^2
\label{eq:nonlinear}
\end{equation}

We estimate $\{x,\alpha\}$ separately for blue and red channels.  We consider only dSph candidate stars with multiple, independent velocity measurements that pass quality control filters (Section \ref{subsec:qc}).  For the blue (red) channel we use 1249 (561) measurements of 583 (257) stars.  We replace the unknown $V_{*j}$ with an estimate computed from the weighted mean, $\hat{V}_{*j}=\sum_{i=1}^{n_j}w_{ij}V_{ij}/\sum_{i=1}^{n_j}w_{ij}$.  For weights we use $w_{ij}=(\sigma_{FX,ij}^2+\sigma_0^2)^{-1}$, where $\sigma_{FX,ij}$ is the initial velocity error returned by FXCOR (to first order, $\sigma_{FX,ij}$ is proportional to $\sigma_{CCF,ij}$).  This substitution introduces a bias, as it ignores the variance in the estimate $\hat{V}_{*j}$.  However, analysis of repeat measurements (Section \ref{subsec:repeats}) demonstrates that resulting error estimates are reasonable. 

\begin{figure}
  \epsscale{1.2}
  \plotone{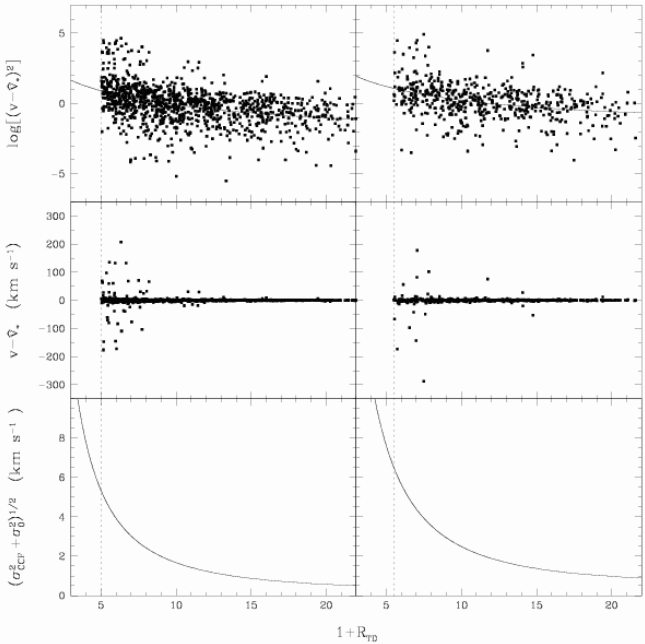}
  \caption{ \textit{Top:} Data from repeat velocity measurements, used for estimating error model parameter pair $\{x,\alpha\}$, with best-fit curve overplotted.  Left/right panels show data for blue/red channels.  \textit{Middle:} Deviation of measured velocity from estimated true velocity.  \textit{Bottom}: Size of error bar resulting from parameter pair $\{x,\alpha\}$ that yields the best-fit curve.  In all panels, dotted, vertical lines indicate the minimally acceptable value of $R_{TD}$ (section \ref{subsec:qc}).}
  \label{fig:xalphafit}
\end{figure}

The top panels in Figure \ref{fig:xalphafit} plot (for blue and red channels) the nonlinear regression data in the $\log[(V_{ij}-\hat{V}_{*j})^2]$ versus $(1+R_{TD,ij})$ plane.  Overplotted is the best-fit curve.  We note the presence of outliers that have $|V_{ij}-\hat{V}_{*j}|$ up to $\sim 250$ km s$^{-1}$ (middle panels of Figure \ref{fig:xalphafit}).  The broad velocity distribution of these outliers suggests an error component not considered in the model specified by Equation \ref{eq:errormodel}.  We address non-Gaussian error in Section \ref{subsec:nongaussianerrors}.

Table \ref{tab:errors} summarizes the estimated parameters of the Gaussian error model.  For the blue (red) channel, the best-fit curve to the regression data corresponds to the parameter pair $\{x,\alpha\}=\{1.69,81.0$ km s$^{-1}\}$ ($\{1.65,107.9$ km s$^{-1}\}$).  Using these values, we calculate error bars $\pm \sigma_{V,ij}$ as a function of $R_{TD}$ (bottom panels of Figure \ref{fig:xalphafit}).  Velocities measured from minimally acceptable CCFs (those with $R_{TD}=R_{TD,c}$), have $\sigma_{V,ij}=5.3$ km s$^{-1}$ (blue) and $\sigma_{V,ij}=6.5$ km s$^{-1}$ (red).  Among all acceptable measurements, median errors are $\sigma_{V,ij}=1.6$ km s$^{-1}$ (blue) and $\sigma_{V,ij}=2.4$ km s$^{-1}$ (red).  
 
\renewcommand{\arraystretch}{0.6}
\begin{deluxetable*}{lrrrrrrr}
  \tabletypesize{\scriptsize}
  \tablewidth{0pc}
  \tablecaption{ Summary of Parameters for Gaussian Error Model}
  \tablehead{\colhead{Channel}&R$_{TD,c}$&\colhead{$N_{repeats}$}&\colhead{$\sigma_0$}&\colhead{$\hat{x}$}&\colhead{$\hat{\alpha}$}&\colhead{$\sigma_{V,median}$}&\colhead{$\sigma_{V,max}$}\\
    \colhead{}&\colhead{}&\colhead{}&\colhead{(km s$^{-1}$)}&\colhead{}&\colhead{(km s$^{-1}$)}&\colhead{(km s$^{-1}$)}&\colhead{(km s$^{-1}$)}
}
  \startdata
  Blue&4.0&1249&0.26&1.69&81.0&1.6&5.3\\
  Red&4.5&561&0.60&1.65&107.9&2.4&6.5
  \enddata
  \label{tab:errors}
\end{deluxetable*}

\subsection{Non-Gaussian Errors}
\label{subsec:nongaussianerrors}

The adopted error model assumes the measurement error is the sum of Gaussian components.  Figure \ref{fig:xalphafit} shows that the several outliers from the main distribution of regression data (top panels) correspond to outliers in the $V_{ij}-\hat{V}_{*j}$ distribution (middle panels).  The distribution of $V_{ij}-\hat{V}_{*j}$ among these outliers indicates an additional, non-Gaussian error component not considered in Equation 2.  For stars with repeat observations it is straightforward to identify and discard affected measurements; this is not possible for stars with only a single measurement.  However, we show that only a negligible number of measurements affected by large non-Gaussian errors are likely to be included in dSph samples.

Potential sources of non-Gaussian errors include astrophysical velocity variability (e.g., binary stars), but the most plausible origin of large $V_{ij}-\hat{V}_{*j}$ outliers is the selection of a false peak from a noisy CCF.  The outlying data points occur preferentially at small $R_{TD}$, a regime in which CCFs are relatively noisy and odds of selecting a false peak increase.  Adjacent peaks in CCFs measured from our spectra are separated typically by $50-100$ km s$^{-1}$.  Assuming false peaks occur at no preferred velocity within the CCF, velocities derived from false peaks will result in large deviations that follow a uniform velocity distribution.  This is consistent with the observed distribution of $V_{ij}-\hat{V}_{*j}$ among outliers (middle panels in Figure \ref{fig:xalphafit}).

To what extent does false peak selection affect our data set?  Among the dSph candidate stars with multiple measurements, we consider cases of false peak selection to be associated with measurements having $|V_{ij}-\hat{V}_{*j}| > 30$ km s$^{-1}$ and $R_{TD} < 6$; we do not consider $V_{ij}-\hat{V}_{*j}$ outliers with $R_{TD} \geq 6$ to be false-peak measurements because all such cases can be attributed to a poor estimate of $V_{*j}$ due to an associated repeat measurement having $R_{TD} < 6$.  For the blue (red) channel, 22 of 1249 (6 of 561) repeat measurements that pass both of the initial quality control filters (Section \ref{subsec:qc}) meet the false-peak criterion.  Combining results from both channels yields 486 additional repeat measurements that pass the quality control filters; ten of these meet the false peak criterion.  Extrapolating to the entire sample, we expect fewer than $2\%$ of measurements passing quality control filters to be derived from false CCF peaks.  Cases that are identifiable via comparison to repeat measurements are removed from the sample, but we expect $\sim 70$ cases of false peak selection to remain among the 4175 stars for which we obtained only a single measurement.  However, if the distribution of unidentified false-peak velocities is uniform over a range spanning at least 400 km s$^{-1}$, as Figure \ref{fig:xalphafit} suggests, we expect only $10-15\%$ to fall within the range of a typical dSph velocity distribution.  This results in only a handful of false-peak velocities spread among the four dSph member samples.  This level of contamination is below that expected to be introduced by cases of ambiguous dSph membership.  

\subsection{Repeat Measurements}
\label{subsec:repeats}

\begin{figure}
  \epsscale{1.2}
  \plotone{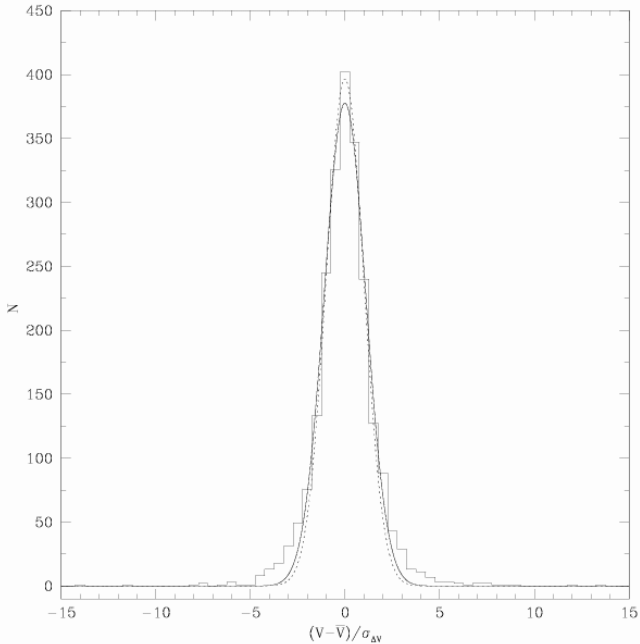}
  \caption{ For all stars with repeat MMFS measurements, the distribution of $V-\bar{V}$, normalized by the estimated error.  Overplotted is the best-fitting Gaussian distribution (solid line), which is nearly identical to the (vertically scaled) normal distribution with unit variance (dotted line).}
  \label{fig:dsph_velrepeats}
\end{figure}

After removing identified cases of false $CCF$ peaks, the final MMFS data set contains repeat velocity measurements for 1011 dSph target stars.  There are 2246 independent measurements of these stars, including up to five measurements for some stars.  If we calculate the weighted mean velocity for the $j^{th}$ star as $\bar{V}_{j}=\frac{\sum_{i=1}^{n_j}(w_{ij}V_{ij})}{\sum_{i=1}^{n_j}w_{ij}}$, using weights $w_{ij}=\sigma_{V,ij}^{-2}$, then the Values $V_{ij}-\bar{V}_j$ should follow a Gaussian distribution with variance 
\begin{equation}
  \sigma_{\Delta V,ij}^2 \equiv  Var\bigl (V_{ij}-\bar{V}_j) =\frac{1}{w_{ij}} \bigl (1-\frac{w_{ij}}{K} \bigr )^2+\frac{1}{K^2}\sum_{k\neq i}^{n_j}w_{kj},
\label{eq:repeatvariance}
\end{equation}
where $K \equiv \Sigma_{i=1}^{n_j} w_{ij}$.  Figure \ref{fig:dsph_velrepeats} plots the distribution of $(V_{ij}- \bar{V}_{j})/\sigma_{\Delta V,ij}$ for all dSph targets with repeat velocity measurements.  A Gaussian fit to the empirical distribution has standard deviation $\sigma=1.28$.  The excess of this value with respect to the nominal value of one is due in large part to the presence of outliers and is not surprising given the likely presence of binaries in the sample.

\section{Comparison with Previous Work}

The MMFS sample has stars in common with several previously published RV data sets---\textbf{Carina}: \citet{mateo93,majewski05,munoz06}.  \textbf{Fornax}: \citet{mateo91,walker06a}.  \textbf{Sculptor}: \citet{armandroff86,queloz95}.  \textbf{Sextans}: \citet{dacosta91,suntzeff93,hargreaves94,kleyna04}.  Among previous studies, only \citet{mateo91}, \citet{queloz95}, and \citet{walker06a} measure RV using the magnesium triplet.  All others use the infrared calcium triplet (CaT) near 8500 \AA.  The MMFS sample offers the first opportunity to compare directly large numbers of velocity measurements obtained using both techniques.   

\begin{figure*}
  \epsscale{1}
  \plotone{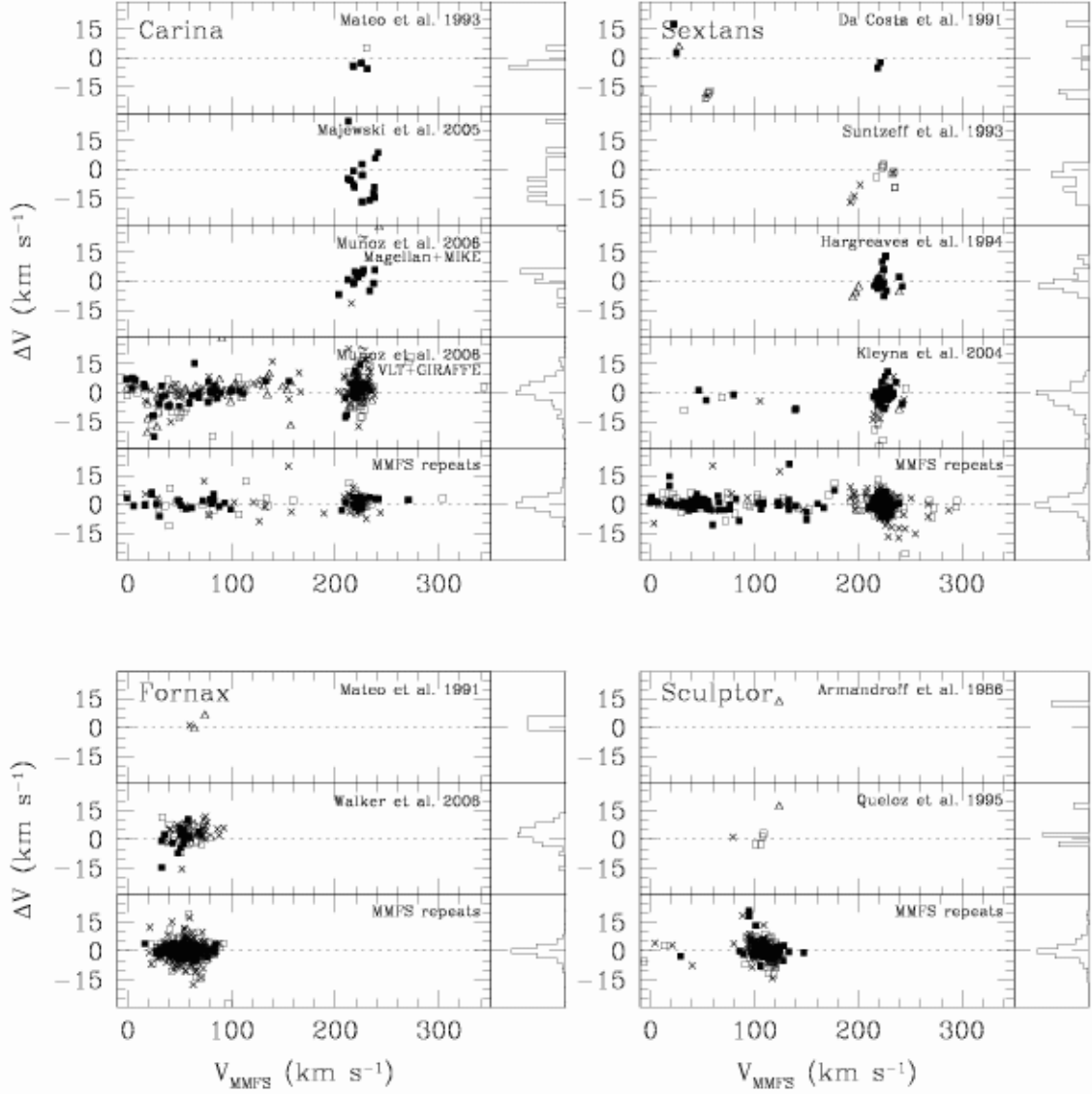}
  \caption{ Comparison of MMFS velocity measurements with previous results.  For stars in common with the specified RV sample, each panel plots the difference in measured velocity ($\Delta V \equiv V_{\rm{MMFS}}-V_{{other}}$) as a function of the MMFS velocity.  The bottom subpanel within each panel plots internal comparisons from stars with multiple MMFS measurements.  Filled squares denote stars for which both the MMFS and comparison measurements have quoted uncertainties less than the median uncertainty in their respective sample.  Open squares/triangles denote stars for which the MMFS measurement has uncertainty less/greater than the median MMFS uncertainty and the comparison measurement has uncertainty larger/less than the median from its sample.  Crosses denote stars for which both the MMFS and published measurements have uncertainties greater than the respective sample median.  Histograms in right-hand sub-panels, individually normalized for legibility, indicate distributions of $\Delta V$.}
  \label{fig:previouswork1}
\end{figure*}

\begin{figure*}
  \epsscale{1}
  \plotone{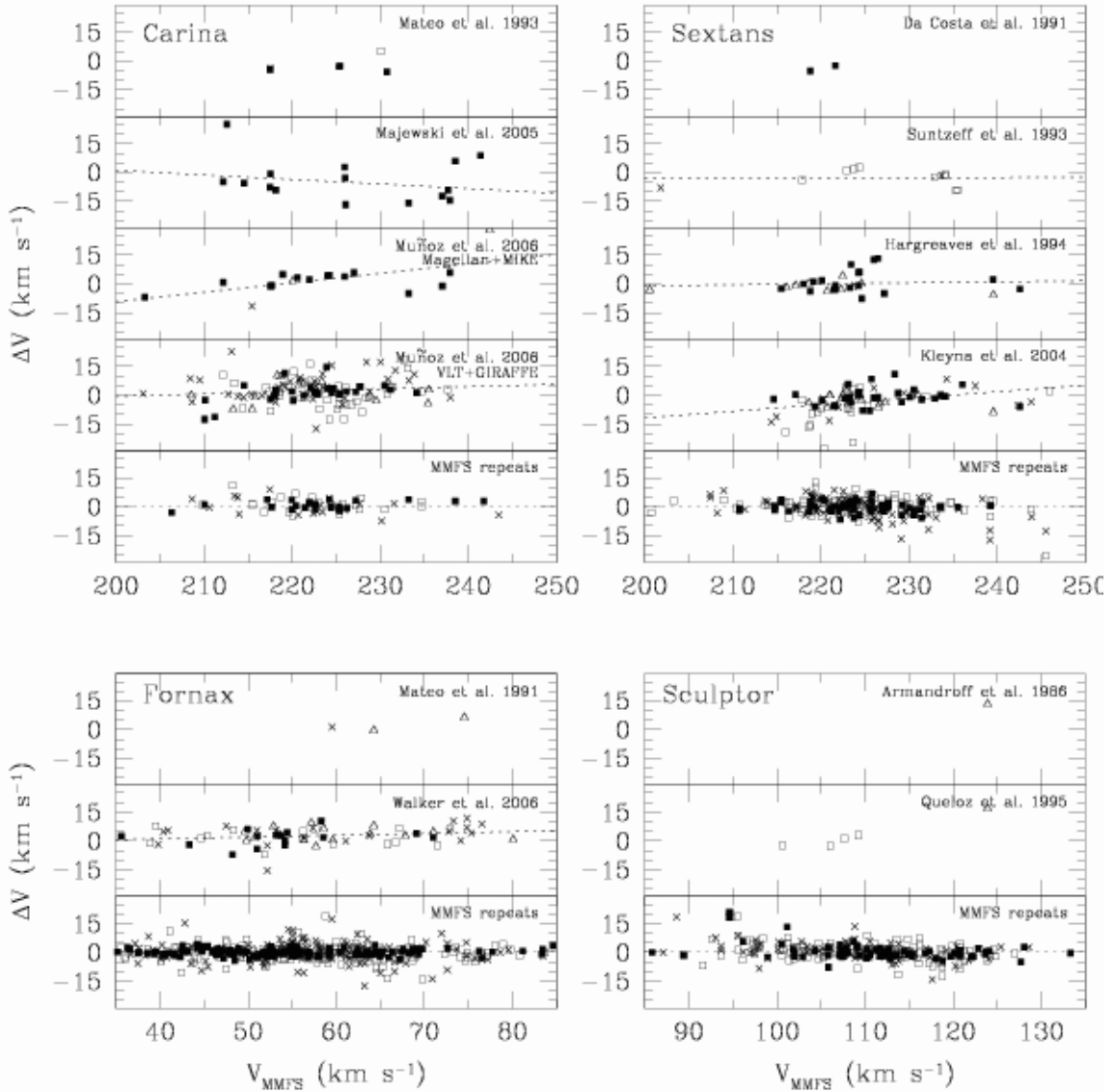}
  \caption{ Comparison of MMFS velocity measurements with previous results, within a window spanning the approximate velocity range of dSph member stars.  Symbols have the same meanings as in Figure \ref{fig:previouswork1}.  The dotted line in each panel provides the best fit to the plotted points.  The steepness of the slopes in the Kleyna et al.\ (2004) and Mu\~{n}oz et al.\ (2006 w/ MIKE) comparisons are significant at the 99.8\% and 98.5\% levels, respectively.}
  \label{fig:previouswork2}
\end{figure*}

\begin{figure*}
  \epsscale{1}
  \plotone{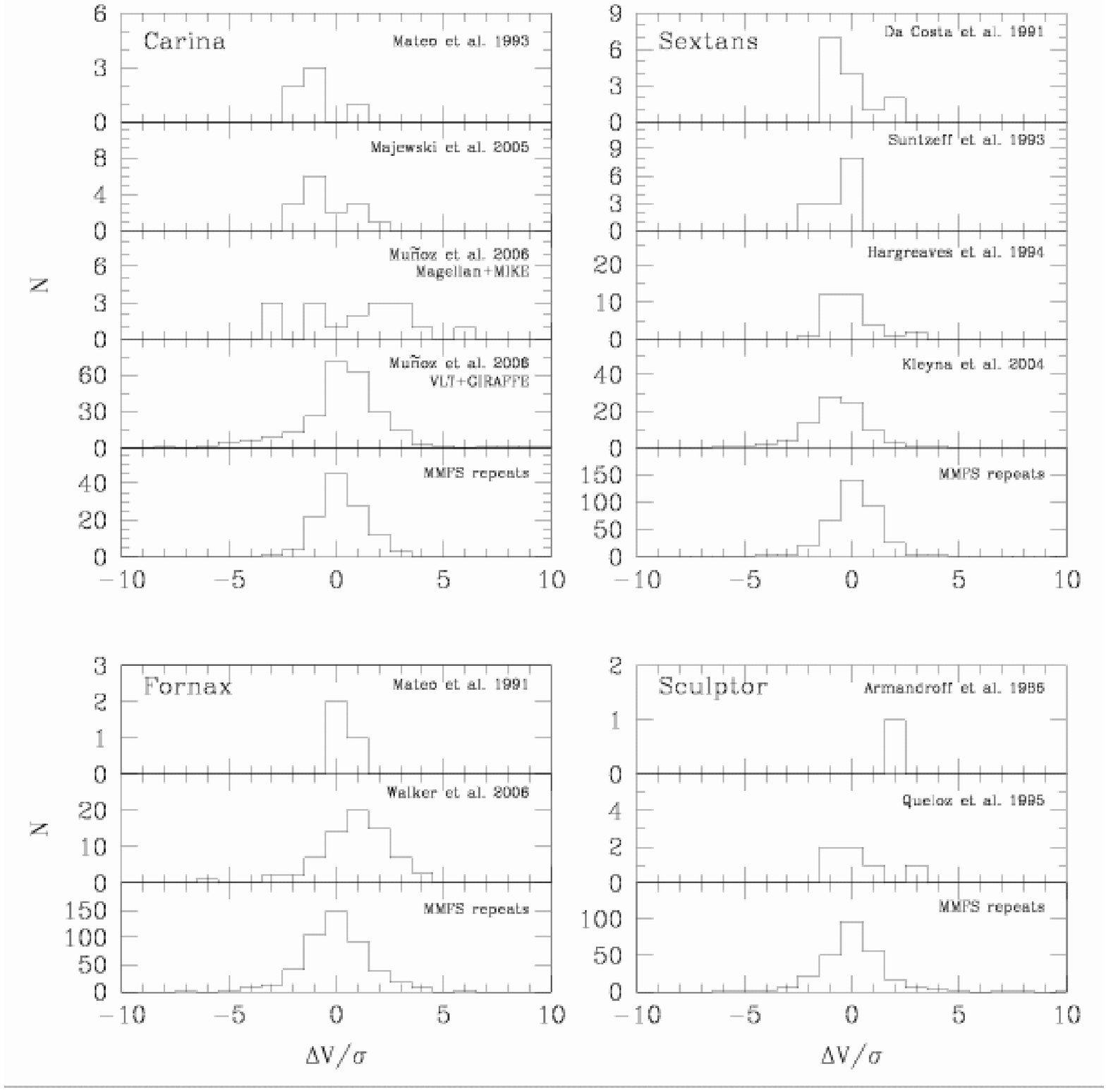}
  \caption{ For MMFS stars in common with previously published samples, distribution of velocity deviations $\Delta V \equiv V_{\rm{MMFS}}-V_{{other}}$ normalized by combined errors $\sigma=\sqrt{\sigma_{MMFS}^2+\sigma_{other}^2}$.}
  \label{fig:previouswork3}
\end{figure*}

To identify MMFS stars in common with other samples, we searched published data sets for coordinate matches, tolerating offsets of up to $0.2\arcsec - 1.5\arcsec$ depending on the sample.  For each detected matching pair, Figure \ref{fig:previouswork1} plots the velocity deviation $\Delta V \equiv V_{\rm{MMFS}}-V_{\rm{other}}$ as a function of $V_{\rm{MMFS}}$.  Figure \ref{fig:previouswork2} plots the same quantities over a window of width 50 km s$^{-1}$, centered on the systemic velocity of the dSph.  Also shown in both figures are internal comparisons using stars with repeat MMFS measurements.  In each panel of Figures \ref{fig:previouswork1} and \ref{fig:previouswork2} the marker type indicates the relative quality of the RV measurement.  Filled squares denote stars for which both the MMFS and comparison measurements have quoted uncertainties less than the median uncertainty in their respective sample.  Open squares/triangles denote stars for which the MMFS measurement has uncertainty less/greater than the median MMFS uncertainty and the comparison measurement has uncertainty greater/less than the median from its sample.  Crosses denote stars for which both the MMFS and comparison measurements have uncertainties greater than the respective sample median.  We find that the best measurements (solid squares) indeed provide the closest agreement between samples.  Figure \ref{fig:previouswork3} plots for each comparison sample the distribution of $\Delta V/\sqrt{\sigma_{\rm{MMFS}}^2+\sigma_{\rm{other}}^2}$.

Table \ref{tab:comparison} summarizes the comparisons.  For each comparison sample, columns 3-6 list the primary absorption features used, the spectral resolution (if available), the number of stars in common with the MMFS sample, and the number of MMFS measurements for the common stars.  Columns 7-8 list the mean velocity offset and standard deviation calculated from all matching pairs.  Because these values are susceptible to the effects of even a single outlier (which may be due to a mis-identified match, a binary star, or a false CCF peak), columns 9-10 list biweight estimates \citep{beers90} of the mean offset and scatter.  While the biweight is resistant to outliers, it is known to perform poorly when the number of matches is small ($N < 10$).  
\renewcommand{\arraystretch}{0.6}
\begin{deluxetable*}{llllrrrrrrrr}
  \tabletypesize{\scriptsize}
  \tablewidth{0pc}
  \tablecaption{ Comparison of RV data with previous results}
  \tablehead{\colhead{Galaxy}&\colhead{Author}&\colhead{$\lambda$}&\colhead{Resolution}&\colhead{$N_*$}&\colhead{$N_{V}$}&\colhead{$\langle \Delta V \rangle$}&\colhead{$\sigma_{\rm{rms}}$}&\colhead{$\langle \Delta V \rangle_{\rm{BW}}$}&\colhead{$\sigma_{\rm{BW}}$}&\colhead{slope}&\colhead{p}\\
      \colhead{}&\colhead{}&\colhead{}&\colhead{}&\colhead{}&\colhead{}&\colhead{(km s$^{-1}$)}&\colhead{(km s$^{-1}$)}&\colhead{(km s$^{-1}$)}&\colhead{(km s$^{-1}$)}&\colhead{}&\colhead{}
}
  \startdata
Carina&\citet{mateo93}&MgT&21000&5&6& -2.34&    3.54&   -0.45&   3.75&     0.27&   0.31200\\
Carina&\citet{majewski05}&CaT&2600,7600\tablenotemark{a}&17& 20&    40.54&  123.32&   -2.04&   12.51&    -0.23&   0.39328\\ 
Carina&\citet{munoz06}\tablenotemark{b}&CaT&19000&11&17&     2.28&    7.94&    0.62&    5.06&     0.49&   0.01460\\
Carina&\citet{munoz06}\tablenotemark{c}&CaT&6500&218&252&    -0.98&   26.88&    0.51&    6.05&     0.13&   0.15372\\
Sextans&\citet{dacosta91}&CaT&2600&6&14&    -5.58&   14.88&   -2.41&   20.71&     1.15&   0.08198\\
Sextans&\citet{suntzeff93}&CaT&2800&4&14&    -5.53&    6.35&   -1.86&    7.50&     0.018&   0.88844\\
Sextans&\citet{hargreaves94}&CaT&---&11&32&     -0.42&    5.26&   -0.61&    4.64&     0.050&   0.68364\\
Sextans&\citet{kleyna04}&CaT&---&45&92&    -6.84&   24.80&   -0.80&    5.46&     0.34&   0.00178\\
Fornax&\citet{mateo91}&MgT&21000&3&3&     2.41&    2.82&    0.40&   1.40&     0.37&   0.14845\\
Fornax&\citet{walker06a}&MgT&21000&58&72&     3.33&   7.76&    1.62&    5.23&    0.094&   0.07913 \\
Sculptor&\citet{armandroff86}&CaT&---&1&1&13.32&\nodata&\nodata&\nodata&\nodata&\nodata\\
Sculptor&\citet{queloz95}&MgT&16000&5&6&     2.94&    6.56 &  0.42 &   2.62 &    0.90&   0.02979\\
\enddata
\tablenotetext{a}{Instrumental setup differed over two observing runs; see \citet{majewski05} for details.}
\tablenotetext{b}{using Magellan+MIKE}
\tablenotetext{c}{using VLT+GIRAFFE}
  \label{tab:comparison}
\end{deluxetable*}

We find generally acceptable agreement with the results of \citet{dacosta91,mateo91,mateo93,suntzeff93,hargreaves94,queloz95,majewski05}.  Despite large apparent deviations, we include the \citet{dacosta91} and \citet{majewski05} samples in this list because both studies report large measurement uncertainties (10-20 km s$^{-1}$ and 4-10 km s$^{-1}$, respectively).  The somewhat large offset and standard deviation with respect to the \citet{queloz95} Sculptor sample are due to a single star; the remaining five matching stars show excellent velocity agreement.  The single MMFS star in common with the \citet{armandroff86} Sculptor sample happens to be the same star giving the most deviant result in the Queloz et al.\ comparison; notice that for this star the Armandroff et al.\ and Queloz et al.\ measurements agree.  Unless the MMFS measurement is in error, it is likely that this star is a binary, consistent with the conclusion of Queloz et al.\ that Sculptor has a binary fraction of $\sim 20\%$.  Other than this, we can conclude little from the single star in common with the \citet{armandroff86} sample.  Similarly, the three stars in common with the \citet{mateo91} Fornax sample offer little basis for comparison.  

The comparison to the Fornax sample of \citet{walker06a} shows evidence for a zero-point offset of $\sim 2$ km s$^{-1}$, and the scatter is somewhat larger than expected from quoted measurement errors.  Here we can rule out mis-identified coordinate matches, as MMFS targets are chosen from the same target list used by \citet{walker06a}.  While it remains possible that some disagreement is due to binary stars, we note that the most deviant measurements correspond to stars for which the \citet{walker06a} measurement has higher uncertainty than the median for that sample.  If the measurement errors for the \citet{walker06a} stars are 1.3 times larger than quoted, the distribution of $\Delta V$ becomes that which we expect for Gaussian errors.  We note that the error model given by Equation \ref{eq:errormodel} improves upon that used in \citet{walker06a}, which does not take the baseline variance into account.  

The \citet{kleyna04} and \citet{munoz06} samples offer relatively large numbers of common stars (45 and 229, respectively) with which to compare MMFS velocities.  The Mu\~{n}oz et al.\ sample consists of two data sets acquired independently using Magellan+MIKE (in slit mode; 11 MMFS stars) and VLT+GIRAFFE (fibers; 218 MMFS stars); we consider each sample separately.  The most striking feature we find in these comparisons is the apparent correlation of $\Delta V$ with $V_{\rm{MMFS}}$.  With respect to the VLT+GIRAFFE data, this manifests as a tendency toward ($V_{\rm{MMFS}}-V_{\rm{VLT}})<0$ km s$^{-1}$ when $V_{\rm{MMFS}}< 75$ km s$^{-1}$ (Figure \ref{fig:previouswork1}).  We note that the stars with largest VLT uncertainties and smallest MMFS uncertainties (open squares) are most responsible for this trend.  Moreover, for the several stars that have incompatible MMFS and VLT velocities \textit{and} have multiple MMFS measurements, we find that the MMFS measurements show good internal agreement.  

In comparison to the Mu\~{n}oz et al.\ MIKE data for Carina and the \citet{kleyna04} data for Sextans, we find apparent correlations between $\Delta V$ and $V_{\rm{MMFS}}$ over the velocity distribution characteristic of the dSph member stars.  Dotted lines in each panel of Figure \ref{fig:previouswork2} give least-squares linear fits to the plotted points.  Columns 11-12 in Table \ref{tab:comparison} give the slope of each line and the probability that such a slope would be observed in the absence of a correlation (determined via 100000 Monte Carlo realizations that assume the same $V_{\rm{MMFS}}$ values and draw $\Delta V$ randomly from a Gaussian distribution with variance equal to the standard deviation of $\Delta V$ for points plotted in Figure \ref{fig:previouswork2}).  The apparent trends observed with respect to the Mu\~{n}oz et al.\ (MIKE) and Kleyna et al.\ comparisons have slopes $\rm{d}(\Delta V)/\rm{d}V_{\rm{MMFS}}=0.49$ and $0.34$, respectively, and these are significant at the 98.5\% 99.8\% levels.  Over the range of velocities consistent with dSph membership, these slopes amount to systematic deviations of up to 10 km s$^{-1}$.

That we observe such behavior in the largest available comparison samples suggests there may be systematic differences between survey data.  At this point we can only speculate regarding the source of such a systematic trend and whether it might originate in the MMFS or the two previously published samples.  Perhaps the most significant observational difference between ours and the other large dSph surveys is that we observe at the MgT rather than the CaT.  Whereas late-type stars are brighter in the infrared, the primary benefit we enjoy by using the MgT is the relatively clean sky.  In contrast, the CaT region is prone to contamination by atmospheric -OH emission lines (see \citealt{osterbrock96}).  At redshifts characteristic of Carina and Sextans ($\sim 225$ km s$^{-1}$), strong sky lines fall directly on top of lines 1 and 2 of the CaT.  \citet{kleyna04} note that, when using CaT lines individually to measure velocities, lines 1 and 2 tend to give spurious Sextans-like velocities of 233 and 223 km s$^{-1}$, respectively, due to contribution from sky residuals to the cross-correlation function (see their Figure 1).  In the end, \citet{kleyna04} use lines 2 and 3 jointly for their velocity measurements.  If the inclusion of line 2 tends to ``pull'' their velocities toward 223 km s$^{-1}$, this could account for the observed gradient in Figure \ref{fig:previouswork2}.  Mu\~{n}oz et al.\ (w/ MIKE) measure Carina velocities using $\sim 15$ absorption features, including all three CaT lines.  Because Carina's radial velocity in the solar rest frame is nearly identical to that of Sextans, one might expect a similar effect from sky residuals.  Indeed we observe in comparison to the Mu\~{n}oz et al.\ (w/MIKE) sample a slope similar in magnitude and significance to those obtained in the comparison to the Kleyna et al.\ sample.  

\section{Spectral Line Indices}

Our MMFS observing program was designed chiefly to measure stellar velocities at low S/N.  While our spectra are not suitable for measuring equivalent widths, nearly all are of sufficient quality for estimating line strengths via spectral indices.  Let $S(\lambda)$ represent the observed spectrum and suppose $\lambda_{f_1}$ and $\lambda_{f_2}$ are lower and upper boundaries, respectively, of a bandpass containing an absorption feature of interest.  A ``pseudo-'' equivalent width is given by the spectral line index defined, for example, by Equation A2 of Cenarro et al.\ (2001; see also \citealt{gonzalez93,cardiel98}):
\begin{equation}
W(\mathrm{\AA})\equiv \int_{\lambda_{f_{1}}}^{\lambda_{f_{2}}}[1-S(\lambda)/C(\lambda)]d\lambda,
\label{eq:index}
\end{equation}
where $C(\lambda)$ is is an estimate of the continuum flux.  Letting $\lambda_{r_1}$,$\lambda_{r_2}$ and $\lambda_{b_1}$,$\lambda_{b_2}$ denote boundaries of nearby continuum regions redward and blueward, respectively, of the feature bandpass, $C(\lambda)$ is given by \citet{cenarro01}:
\begin{equation}
C(\lambda) \equiv S_b \frac{\lambda_r-\lambda}{\lambda_r-\lambda_b}+S_r \frac{\lambda-\lambda_b}{\lambda_r-\lambda_b}.
\end{equation} 
Here $\lambda_r$ and $\lambda_b$ are the central wavelengths of the red and blue continuum bandpasses, respectively, and $S_b$ and $S_r$ are estimates of the mean level in the two continuum bands: 
\begin{eqnarray*}
S_b \equiv (\lambda_{b_{2}}-\lambda_{b_{1}})^{-1}\int_{\lambda_{b_{1}}}^{\lambda_{b_{2}}}S(\lambda)\rm{d}\lambda;\\
S_r \equiv (\lambda_{r_{2}}-\lambda_{r_{1}})^{-1}\int_{\lambda_{r_{1}}}^{\lambda_{r_{2}}}S(\lambda)\rm{d}\lambda.
\label{eq:sbsr}
\end{eqnarray*}

The Lick system of indices \citep{burstein84,faber85,worthey94}, while in wide use, is not applicable to our spectra, as the relevant Lick magnesium passband alone is larger than our entire spectral range!  Instead we take advantage of the high MMFS spectral resolution and define a set of sixteen indices, each of which measures the flux in a single, resolved spectral line.  

\begin{figure*}
  \plotone{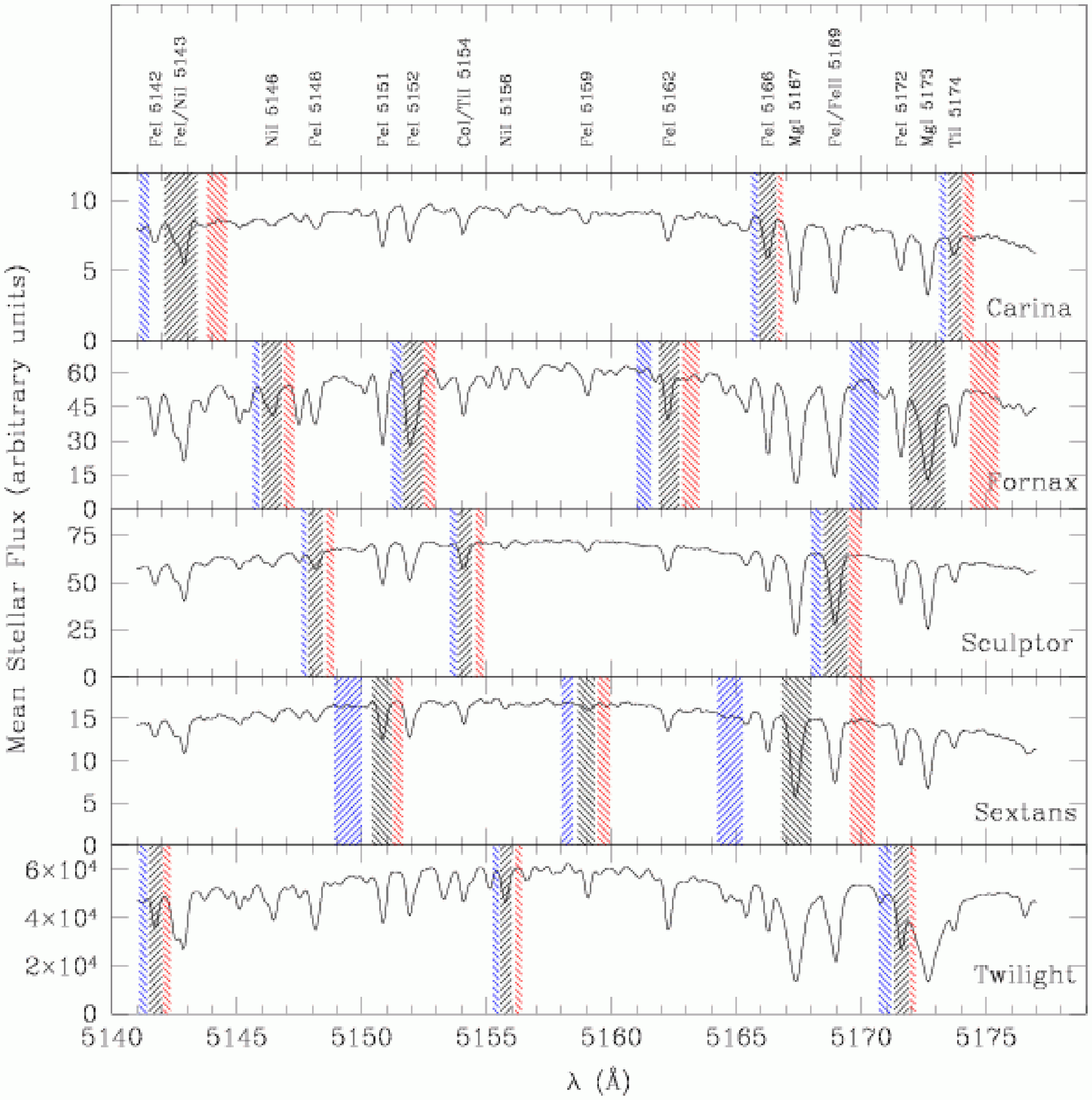}
  \caption{ Composite spectra produced by averaging (with redshifts and spectra from probable nonmembers removed) individual dSph stellar and twilight spectra acquired with MIKE's blue channel.  Black shaded regions correspond to feature bandpasses for sixteen MMFS line indices.  Blue and red shaded regions correspond to the associated continuum bandpasses.  To avoid overlap, each panel shows a unique subset of the sixteen indices.
}
  \label{fig:specbandsblue}
\end{figure*}

\begin{figure*}
  \plotone{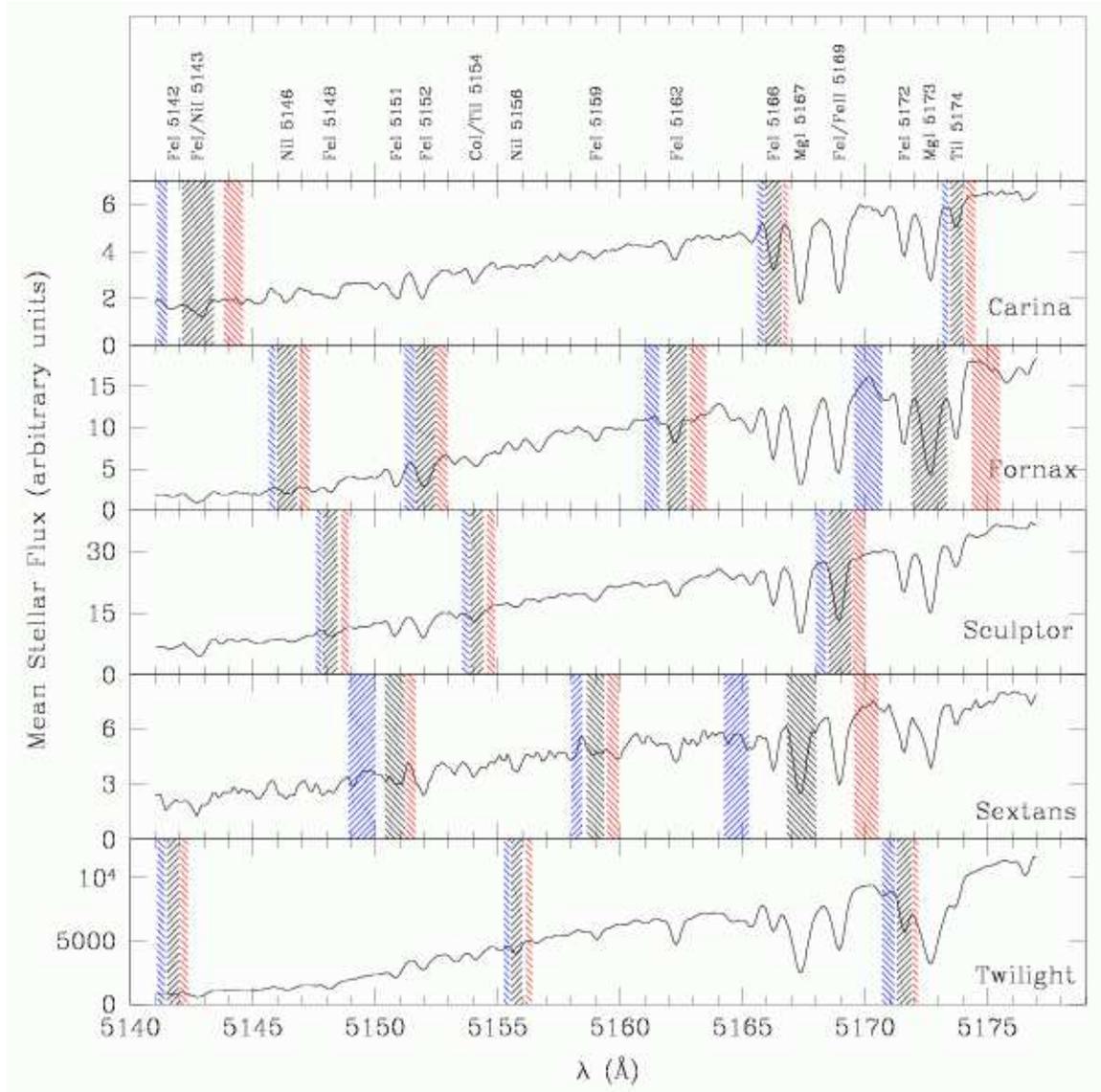}
  \caption{ Same as Figure \ref{fig:specbandsblue} but for spectra acquired with MIKE's red channel.}
  \label{fig:specbandsred}
\end{figure*}

To aid in the determination of feature and continuum bandpasses appropriate for the MMFS spectra we produced a high-S/N composite spectrum for each galaxy by averaging its individual stellar spectra (with relative redshifts and probable nonmembers removed).  Figures \ref{fig:specbandsblue} and \ref{fig:specbandsred} display composite dSph spectra for blue and red channels.  We define feature bandpasses to cover the sixteen atomic absorption lines most prominent in these spectra.  In practice the width of each feature bandpass is approximately proportional to the mean line strength in the composite spectra.  In most cases, blue and red continuum bandpasses are adjacent to the feature bandpass and span at least 0.25 \AA\ ($\sim 5$ pixels).  Where possible (in the absence of a nearby feature) we use larger continuum passbands.  Feature and continuum passbands for the chosen indices appear as shaded regions over the blue (red) composite galaxy spectra in Figure \ref{fig:specbandsblue} (Figure \ref{fig:specbandsred}).  Table \ref{tab:indexbands} lists bandpass boundaries for the sixteen MMFS indices.


\begin{deluxetable*}{lccc}
  \tabletypesize{\scriptsize}
  \tablewidth{0pc}
  \tablecaption{ MMFS Spectral Line Indices}
  \tablehead{\colhead{Index}&\colhead{Central Bandpass}&\colhead{Red Continuum}&\colhead{Blue Continuum}\\
    &\colhead{(\AA)}&\colhead{(\AA)}&\colhead{(\AA)}
  }
  \startdata
FeI$_{5142}$&$5141.50-5142.00$&$ 5141.10-5141.40$&$ 5142.05-5142.35$\\
FeI/NiI$_{5143}$&$5142.10-5143.40$&$5141.10-5141.50$&$5143.80-5144.60$\\
NiI$_{5146}$&$5146.00-5146.80$&$ 5145.65-5145.90$&$ 5146.90-5147.30$\\
FeI$_{5148}$&$5147.85-5148.45$&$ 5147.60-5147.80$&$ 5148.60-5148.90$\\
FeI$_{5151}$&$5150.40-5151.20$&$ 5148.90-5150.00$&$ 5151.25-5151.65$\\
FeI$_{5152}$&$5151.65-5152.45$&$ 5151.20-5151.60$&$ 5152.50-5152.95$\\
CoI/TiII$_{5154}$&$5153.85-5154.40$&$5153.55-5153.80$&$5154.55-5154.90$\\
NiI$_{5156}$&$5155.55-5156.00$&$ 5155.25-5155.50$&$ 5156.15-5156.40$\\
FeI$_{5159}$&$5158.65-5159.35$&$ 5158.00-5158.45$&$ 5159.45-5159.95$\\
FeI$_{5162}$&$5161.90-5162.70$&$ 5161.00-5161.60$&$ 5162.85-5163.50$\\
FeI$_{5166}$&$5165.90-5166.60$&$ 5165.60-5165.85$&$ 5166.65-5166.85$\\
MgI$_{5167}$&$5166.80-5168.00$&$ 5164.25-5165.25$&$ 5169.55-5170.70$\\
FeI/FeII$_{5169}$&$5168.50-5169.45$&$5168.00-5168.40$&$5169.50-5170.00$\\
FeI$_{5172}$&$5171.30-5171.90$&$ 5170.70-5171.20$&$ 5171.95-5172.15$\\
MgI$_{5173}$&$5171.90-5173.35$&$ 5169.55-5170.70$&$ 5174.35-5175.50$\\
TiI$_{5174}$&$5173.50-5174.00$&$ 5173.15-5173.40$&$ 5174.10-5174.50$\\
\enddata
  \label{tab:indexbands}
\end{deluxetable*}  

\subsection{Measurement of MMFS Indices}

Further processing of individual stellar spectra is required prior to measuring line indices.  For dSph and globular cluster stellar spectra, we begin with the sky-subtracted frames obtained during the data reduction steps described in Section \ref{sec:reduction}.  We use the DOPCOR task and the measured radial velocities to shift wavelength solutions such that all spectra have zero redshift in the solar rest frame.  We then truncate spectra such that all cover the range $5141 - 5177$ \AA\ with linear dispersion 0.071 \AA\ pix$^{-1}$ (blue) and 0.107 \AA\ pix$^{-1}$ (red).  These choices of dispersion values minimize pixel re-binning.  From the resulting spectra, which share a common redshift, range, and dispersion, we measure the sixteen MMFS indices using Equations \ref{eq:index}-\ref{eq:sbsr}.  In order to estimate index measurement errors, we assume the amounts by which the measured values $W$ differ from ``true'' index values $W_{*}$ follow Gaussian distributions with variances $\sigma_{W}^2$.  For the $i^{th}$ measurement of the $j^{th}$ star,
\begin{equation}
  W_{ij}-W_{*j}=\sigma_{W,ij}\epsilon_{ij}.
  \label{eq:inderror}
\end{equation}
As in Equation \ref{eq:errormodel}, the values $\epsilon_{ij}$ follow the standard normal distribution, and we adopt the error approximation given by \citet{cardiel98}, 
\begin{equation}
  \sigma_{W} \approx \beta \frac{c_1-c_2W}{\langle SN \rangle},
\label{eq:indexerror}
\end{equation}
modified by our introduction of scaling parameter $\beta$.  Constants $c_1$ and $c_2$ depend on the index definition and are given by Equations 43-44 of \citet{cardiel98}, and we use our Equation \ref{eq:snratio} to estimate $\langle SN \rangle$, the mean S/N per pixel.  After taking logarithms and replacing $\log [\epsilon_{ij}^2]$ with $\delta_{ij}-0.55$ (Section \ref{subsec:gaussianerrors}), we obtain $\log[(W_{ij}-W_{*j})^2]-\log (S_{ij}^2)=\log (\hat{\beta}^2)-0.55$.  We estimate $\beta$ uniquely for each of the sixteen indices on both channels via linear regression, using the approximation $W_{*j} \sim \sum_{i=1}^{n_j}W_{ij}w_{ij}/\sum_{i=1}^{n_j}w_{ij}$ (where $w_{ij}$ is the inverse of the variance obtained from Equation \ref{eq:indexerror} with $\beta=1$).  For the sixteen blue indices, estimates all fall within the range $3.45 \leq \hat{\beta} \leq 4.89$.  For the eight red indices, $2.72 \leq \hat{\beta} \leq 3.95$.  Among all indices, the resulting errors have median values in the range $0.05 \leq \sigma_{W}/\rm{\AA} \leq 0.10$.

\subsection{Individual Index Results}

\begin{figure*}
  \plotone{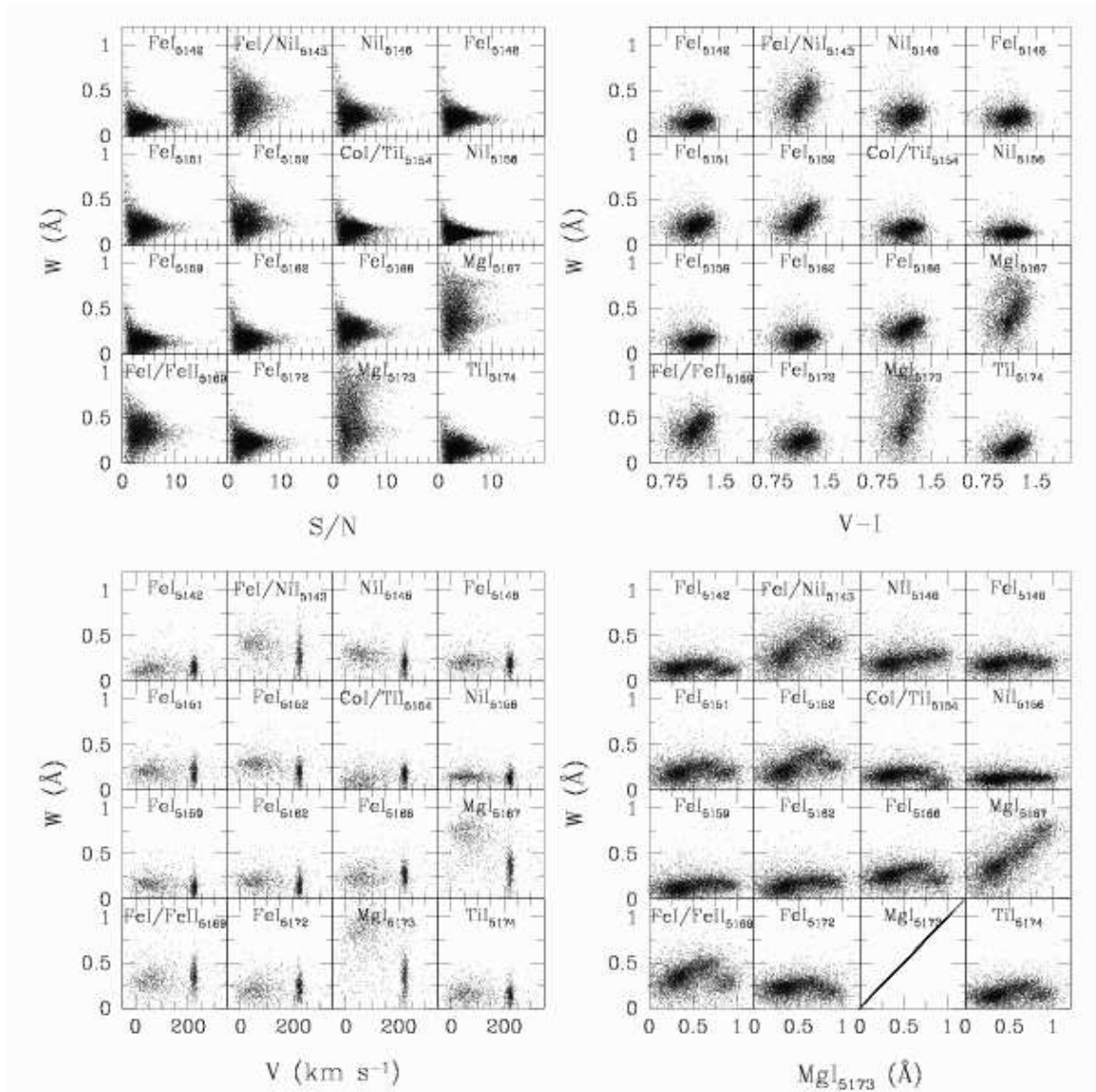}
  \caption{ Measured index value vs.\ spectral and stellar parameters for blue-channel spectra. Each panel corresponds to one of the sixteen MMFS indices.  Points in the lower-left panel correspond to spectra from only Carina target stars (interlopers included), and show a clear separation in MgI indices between dSph and interloper populations.}  
  \label{fig:indicesblue}
\end{figure*}

\begin{figure*}
  \plotone{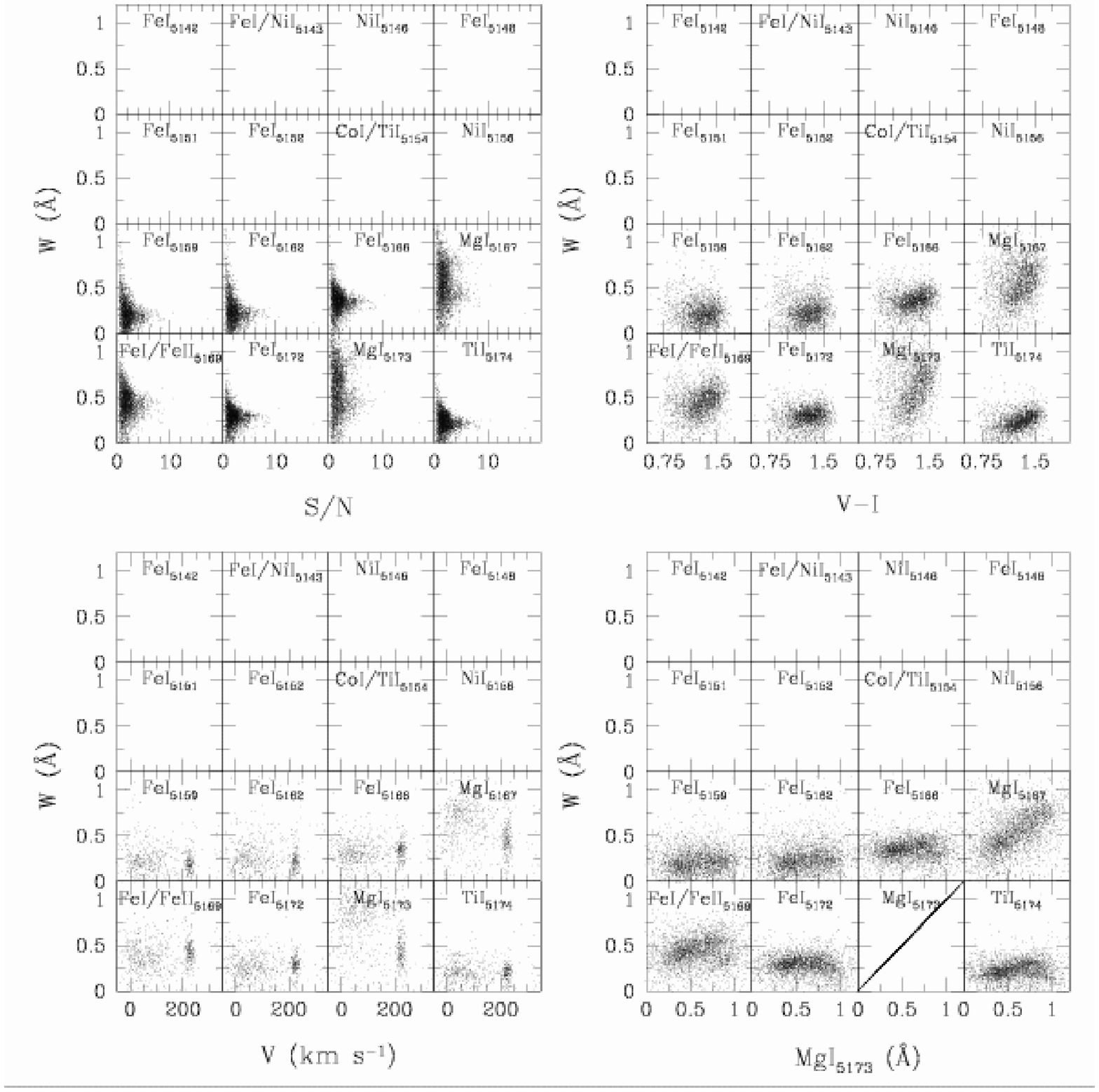}
  \caption{ Same as Figure \ref{fig:indicesblue} but for red-channel spectra.  }
  \label{fig:indicesred}
\end{figure*}

The upper-left panels in Figures \ref{fig:indicesblue} and \ref{fig:indicesred} display, for each index, the measured value as a function of the mean S/N per pixel (calculated over the range $5141-5177$ \AA).  Because the red channel has poor sensitivity toward the blue end of our spectral coverage (see the composite red spectra in Figure \ref{fig:specbandsred}), we consider only the eight red indices with feature bandpass redward of 5158 \AA.  For a given index, the width of the distribution of $W$ values correlates with both (S/N) and the width of the feature bandpass

The remaining panels in Figures \ref{fig:indicesblue} and \ref{fig:indicesred} display empirical relationships between individual MMFS indices and other stellar parameters: V-I color, radial velocity, and MgI$_{5173}$ index.  For several indices there is a clear trend toward higher index values at redder colors (upper-right panel); this behavior indicates the expected dependence of line strength on effective temperature.  The lower-left panels plot $W$ against the measured radial velocity; for clarity in these panels the plotted points correspond exclusively to spectra from Carina targets.  Along the radial velocity axis, most points are clumped in relatively narrow distributions about Carina's systemic velocity ($\sim 225$ km s$^{-1}$).  That the velocity outliers with respect to the narrow Carina distribution tend also to be MgI index outliers indicates the ability of the MgI indices to distinguish dSph members from interlopers.  We expect interlopers with large MgI line strengths to be due to Milky Way dwarf stars with high surface gravity.  Thus the spectral indices provide a basis for dwarf/giant discrimination similar to that derived from photometric techniques that use narrow-band filters (e.g., DDO51) sensitive to surface gravity \citep{geisler84,majewski00}.

Finally, when individual indices are plotted against one-another, indices corresponding to the same element(s) display monotonic, linear relationships with slopes approximately equal to the ratio of feature bandpass widths (if indices are stated as a fractions of their bandpass widths, these relations tend toward 1:1).  For example, the lower-right panels of Figures \ref{fig:indicesblue} and \ref{fig:indicesred} plot each index against the MgI$_{5173}$ index.  In this case the MgI$_{5167}$ index exhibits the expected monotonic, linear relation.  Noting the behavior of the FeI indices, it is immediately striking that many exhibit a bifurcation.  With respect to the trend followed by stars with small and intermediate MgI$_{5173}$ index values, stars with the largest MgI$_{5173}$ values have systematically small FeI indices.  Such stars tend to be the outliers in the radial velocity and MgI index joint distributions (lower-left panels); thus the bifurcation is due primarily to the presence of interloping stars, and dSph members generally follow a linear relationship in the iron-magnesium index plane. 

\subsection{Composite Indices}

\begin{figure}
  \epsscale{1.2}
  \plotone{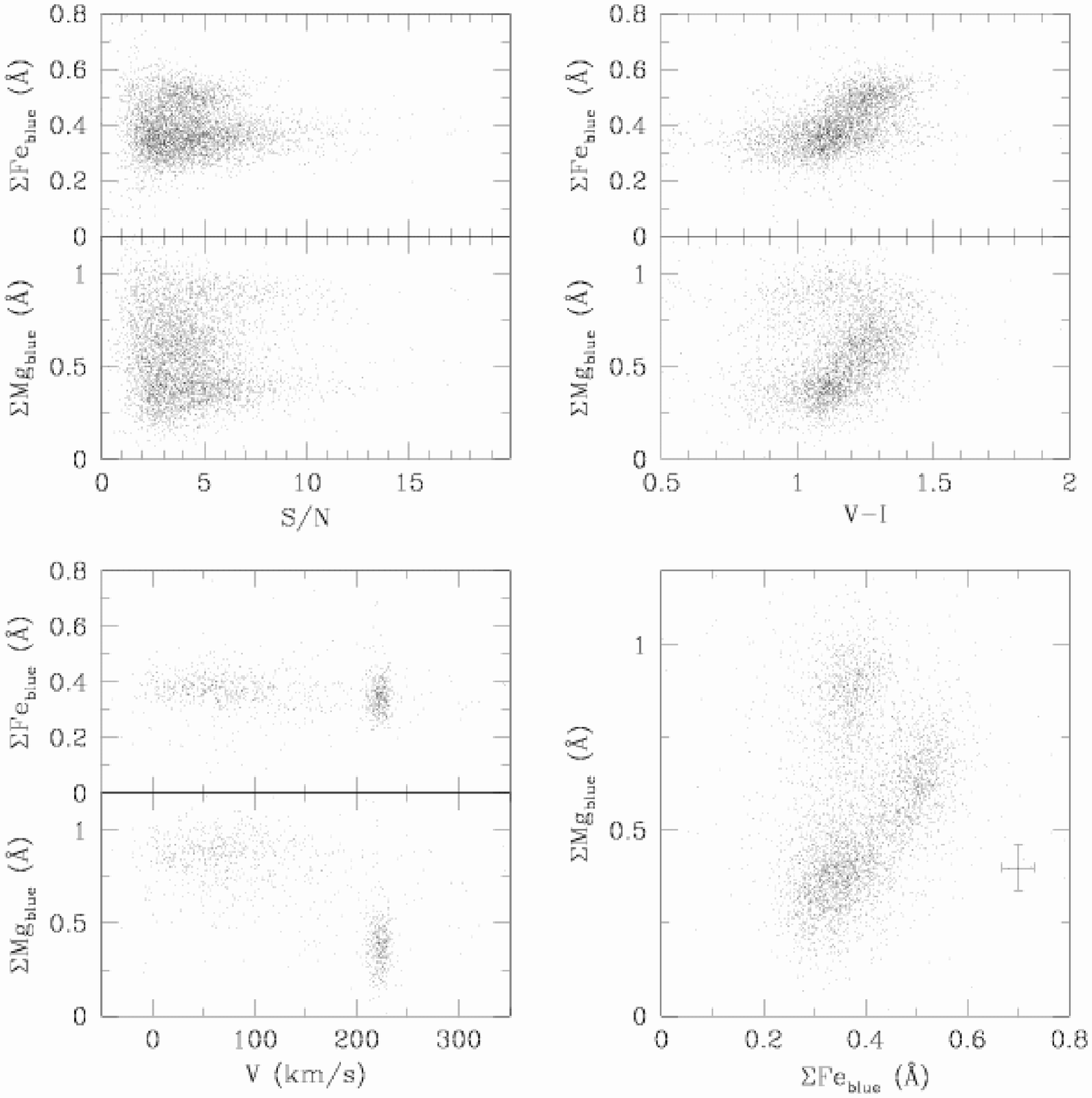}
  \caption{ Dependence of Fe and Mg composite indices on spectral and stellar parameters for blue-channel spectra.  Points correspond to the same spectra plotted in Figure \ref{fig:indicesblue} (again, points in the lower-left panel correspond only to Carina targets).  Errorbars in the lower-right panel indicate median error bars.}
  \label{fig:masterindicesblue}
\end{figure}

\begin{figure}
  \epsscale{1.2}
  \plotone{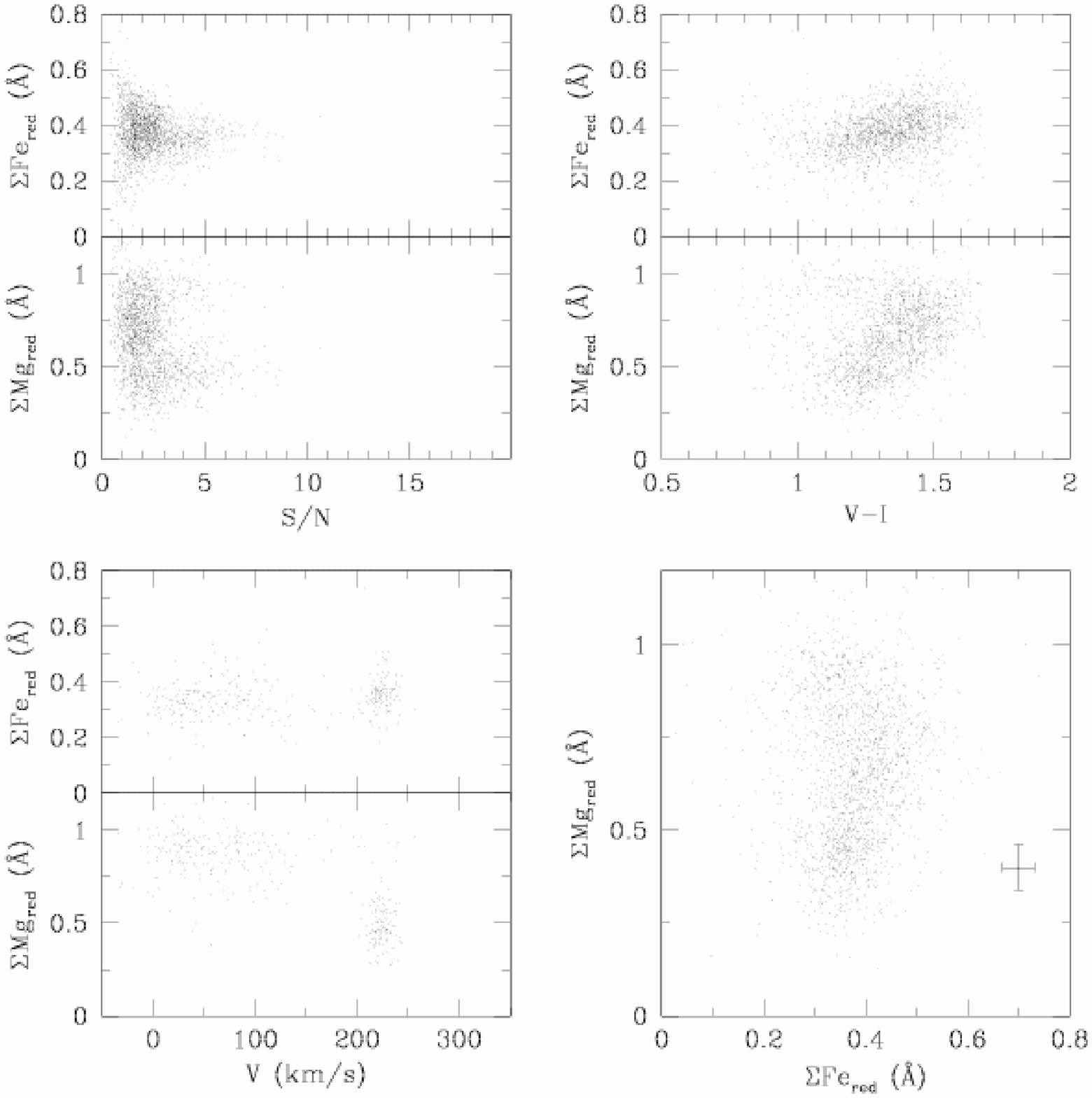}
  \caption{ Same as Figure \ref{fig:masterindicesred} but for red-channel spectra.}
  \label{fig:masterindicesred}
\end{figure}

\begin{figure}
  \epsscale{1.2}
  \plotone{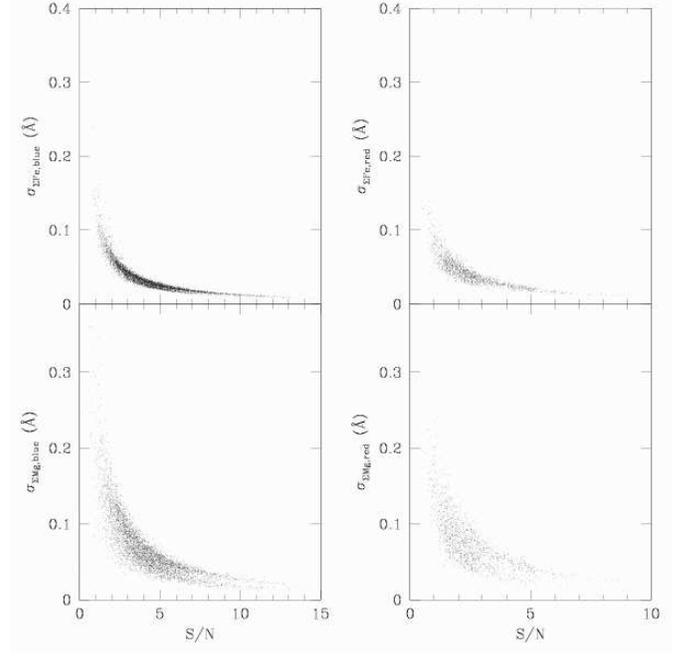}
  \caption{ For composite Fe (top panels) and Mg (bottom panels) indices measured on blue (left) and red (right) channels, sizes of derived error bars are plotted against mean spectral S/N per pixel.  Due to the dependence of $\sigma$ on line strength (Equation \ref{eq:indexerror}), interloping dwarf stars with systematically stronger absorption lines have systematically smaller derived errors, producing the apparent bifurcations in the bottom two panels. }  
  \label{fig:masterindices_errors}
\end{figure}

We have noted that indices for different absorption lines corresponding to the same element(s) correlate linearly.  We can improve precision by computing their sum.  For blue-channel spectra we define a composite Fe index:
\begin{eqnarray}
  \Sigma \rm{Fe}_{blue} \equiv 0.160(\rm{Fe}I_{5142}+\rm{Fe}I/NiI_{5143}+\rm{Fe}I_{5148}\\
  +\rm{Fe}I_{5151}+\rm{Fe}I_{5152}+\rm{Fe}I_{5159}+\rm{Fe}I_{5162}\nonumber\\
  +\rm{Fe}I_{5166}+\rm{Fe}I/\rm{Fe}II_{5169}+\rm{Fe}I_{5172})\nonumber,
  \label{eq:masterfeblue}
\end{eqnarray}
where the sum is scaled by the ratio of the FeI$_{5143}$ bandpass width to the sum of all FeI bandpass widths (this scaling effectively weights each index by its bandpass width).  Similarly, for red-channel spectra the composite Fe index is given by
\begin{eqnarray}
  \Sigma \rm{Fe}_{red} \equiv 0.253(\rm{Fe}I_{5159}+\rm{Fe}I_{5162}+\rm{Fe}I_{5166}\\
  +\rm{Fe}I/\rm{Fe}II_{5169}+\rm{Fe}I_{5172})\nonumber,
  \label{masterfered}
\end{eqnarray}  
where in this case the straight sum is scaled by the ratio of the FeI$_{5169}$ bandpass width to the sum of the included FeI bandpass widths.  
For both blue- and red-channel spectra we compute the composite Mg index as
\begin{equation}
  \Sigma \rm{Mg} \equiv 0.547(\rm{Mg}I_{5167}+\rm{Mg}I_{5173}),
  \label{eq:mastermg}
\end{equation}  
where the scale factor is the ratio of the MgI$_{5173}$ bandpass width to the sum of the two MgI bandpass widths.  We compute errors $\sigma_{\Sigma W}$ associated with composite indices by adding in quadrature the estimated errors $\sigma_{W}$ of the contributing individual indices, and then scaling by the constant multipliers in Equations \ref{eq:masterfeblue}-\ref{eq:mastermg}.  Figures \ref{fig:masterindicesblue} and \ref{fig:masterindicesred} plot for composite indices the same relations shown for individual indices in Figures \ref{fig:indicesblue} and \ref{fig:indicesred}.  The composite indices show the same features and behaviors as the relevant individual indices, only with less scatter.  Figure \ref{fig:masterindices_errors} plots estimated errors as a function of mean S/N per pixel.  

We note that $\Sigma \rm{Fe}_{blue}$ includes a small contribution from a NiI absorption line at 5143 \AA.  This line is blended with two FeI lines in the MMFS spectra.  Because the FeI/NiI$_{5143}$ index behaves similarly to the other FeI indices (see Figures \ref{fig:indicesblue} and \ref{fig:indicesred}) and because it has among the highest S/N of any individual Fe index, we choose to include it in $\Sigma \rm{Fe}_{blue}$.  For similar reasons we choose to include the FeI/FeII$_{5169}$ in both $\Sigma \rm{Fe}_{blue}$ and $\Sigma \rm{Fe}_{red}$ despite a small contribution from a blended FeII feature.  Exclusion of either index from the composite Fe index does not significantly alter any of the correlations or features seen in Figures \ref{fig:masterindicesblue} and \ref{fig:masterindicesred}, nor the distribution of errors seen in Figure \ref{fig:masterindices_errors}.

\begin{figure}
  \epsscale{1.2}
  \plotone{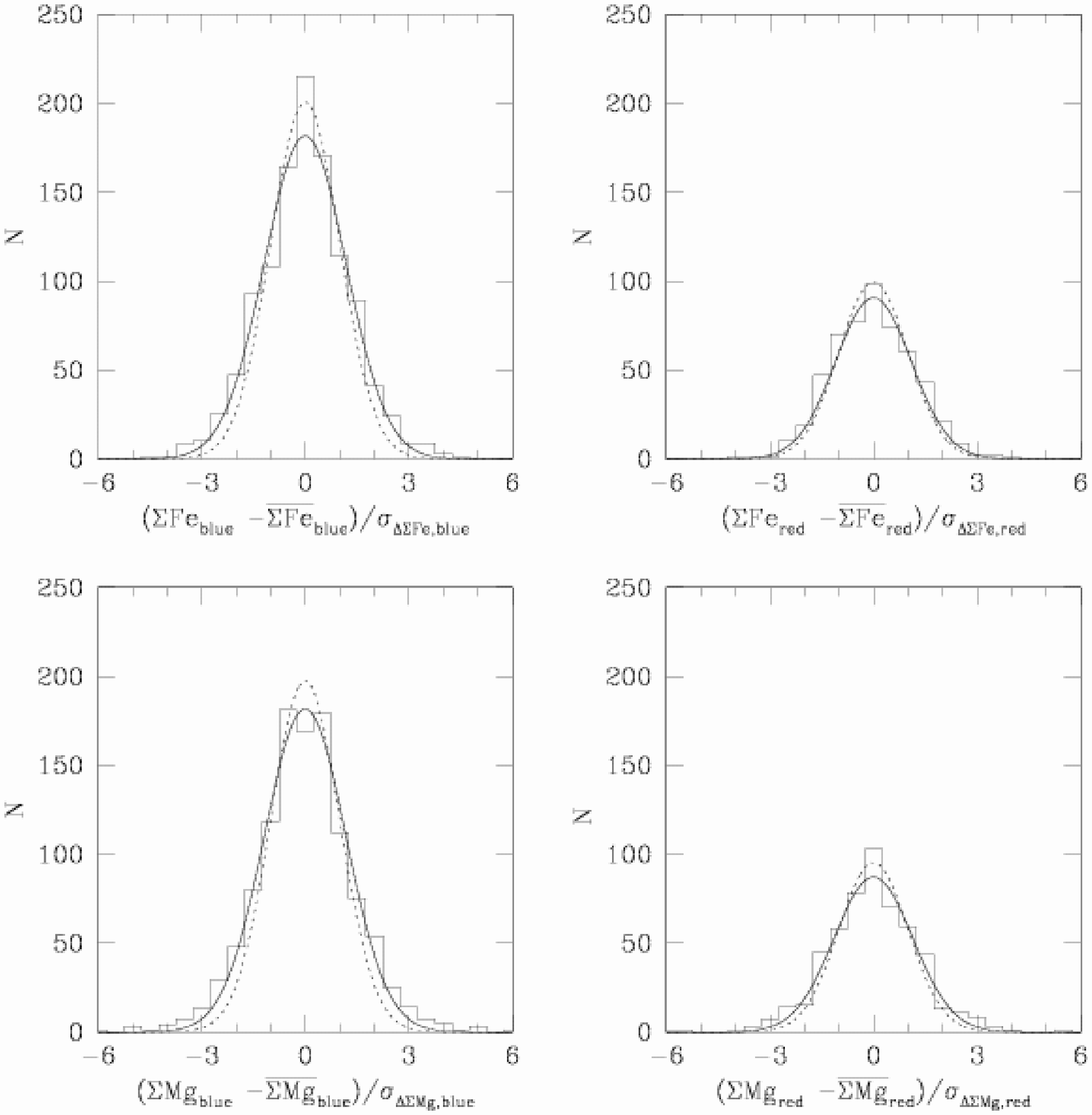}
  \caption{ For stars with repeat index measurements within a given MIKE channel, the distribution of measurements about the weighted mean, normalized by measurement errors.  Solid curves represent the best-fitting Gaussian distributions.  Dotted curves represent a standard normal distribution.}  
  \label{fig:indrepeats}
\end{figure}

For stars with multiple index measurements, Figure \ref{fig:indrepeats} plots the distribution of $\Delta \Sigma W \equiv \Sigma W_{ij} - \bar{\Sigma W_{j}}$, normalized by the associated errors, where $\bar{\Sigma W_{j}}$ is the weighted mean.  The best-fitting Gaussian distributions (solid curves) are nearly identical to the standard normal distribution (dotted curves), indicating that the estimated errors are valid.  

\begin{acknowledgments}

\section{Conclusion}
We have described the instrumentation, data acquisition and data reduction procedures for a new spectroscopic survey that targets large numbers of stars in nearby dSphs.  In forthcoming papers (Walker et al.\ in preparation) we present the entire data set described herein and provide kinematic analyses that consider both equilibrium and tidal interaction models.  We use the spectral indices measured from stars in seven globular clusters to calibrate relations between the MMFS spectral indices and iron abundance, and to identify dSph interlopers independently of velocity.  In the end we have velocity and line-strength measurements for nearly 3800 probable members in the four observed dSphs.  Alone and/or in combination with data published from other contemporary spectroscopic surveys, the MMFS data set enables us to map the two-dimensional behavior of dSph stellar kinematics and chemistry with unprecedented precision. 

We thank Matthew Coleman and Gary Da Costa for providing photometric data that was used to select spectroscopic targets in Sculptor.  We thank Kaspar von Braun and Patrick Seitzer for providing photometric observations that we used to select spectroscopic targets in Carina.  We thank the following people for invaluable assistance in developing MMFS: Stephen Shectman, Steve Gunnels, Alex Athey and Vince Kowal, as well as the excellent staff at Las Campanas Observatory, including Emilio Cerda, Oscar Duhalde, Patricio Jones, Marc Leroy, Gabriel Martin and Mauricio Navarette, and Magellan telescope operators Victor Meri\~{n}o, Hernan Nu\~{n}ez, Hugo Rivera, and Geraldo Valladares.  This work is supported by NSF Grants AST 05-07453, AST 02-06081, and AST 94-13847.
\end{acknowledgments}
\bibliography{ref}

\end{document}